\documentclass[twocolumn]{revtex4-1}

\usepackage{graphicx} 

\begin{document}

\title{Intermittent flow and transient congestions of soft spheres passing narrow orifices}
\author{Kirsten Harth,$^{\ast}$\textit{$^{1}$} Jing Wang,$^{\ast}$\textit{$^{1}$} Tam\'as B\"orzs\"onyi,\textit{$^{2}$}
Ralf Stannarius \textit{$^{1}$} }
\affiliation{$^{1}$Institute of Physics, Otto von Guericke University, Magdeburg, Germany,
$^{2}$Institute for Solid State Physics and Optics, Wigner Research Centre for Physics, Budapest, Hungary,\\
$\ast$shared first authorship}

\begin{abstract}
Soft, low-friction particles in silos show peculiar features during their discharge. The outflow velocity and the clogging
probability both depend upon the momentary silo fill height, in sharp contrast to silos filled with hard particles.
The reason is the fill-height dependence of the pressure at the orifice. We study the statistics of silo discharge of soft hydrogel spheres.
The outflow is found to become increasingly fluctuating and even intermittent with decreasing orifice size, and with decreasing fill height.
In orifices narrower than two particle diameters, outflow can stop completely, but in contrast to clogs formed by rigid particles, these
congestions may dissolve spontaneously. We analyze such non-permanent congestions and attribute them to slow reorganization processes in the container,
caused by viscoelasticity of the material.
\end{abstract}

\maketitle


\section{\label{sec:level1}Introduction}
The peculiarities of granular materials stored in containers have raised the interest of scientists already centuries ago.
One of the earliest scientific reports on the outflow of sand from storage containers dates back to 1829, when Pierre Huber-Burnand
\cite{Huber-Burnand1829} described the pressure conditions in a granular bed within vertical cylinders. He noticed already that the flow of grains is
essentially pressure independent. Gotthilf Hagen (known for the Hagen-Poiseuille law) reported, in great detail, experiments and calculations
on the pressure in dry sand \cite{Hagen1839}. An important step forward was made by Janssen \cite{Janssen1895,Sperl2006} who
measured the pressure in wheat-filled silos and provided a quantitative explanation. He predicted that the pressure characteristics
changes with increasing fill level from a hydrostatic behaviour at fill heights comparable to the container diameter to a saturated maximum pressure.
The latter is fill-level independent, but relates to the friction properties of the grains and the bin width.
Many dynamic features of hard particles discharging from silos with small orifices at the bottom are well-known: The grains flow freely and
without interruptions when the orifice size is sufficiently large (about five particle
diameters or more). The discharge velocity is described quite reliably by Beverloo's equation \cite{Franklin1955,Beverloo1961,Nedderman1982} and
some refinements proposed later \cite{Mankoc2007}.
With smaller orifice sizes, the discharge rate decreases continuously. The outflow rate is practically independent of the
pressure at the container bottom, and thus also independent of the container fill height \cite{Nedderman1982,Anand2008}.
This is true at least when the instantaneous fill height is larger than the orifice size.

Below a certain outlet diameter, hard particles
form a stable clog at the orifice \cite{Beverloo1961,To2001}. This structure blocks further outflow, and it can only be destroyed by
external forcing like vibration of the silo \cite{Wassgren1999,Wassgren2002,Mankoc2009,Janda2009,Lozano2012,Zuriguel2017,Guerrero2018,Guerrero2019}
or air flushes through the opening
\cite{Zuriguel2005}.
The destruction of clogged states by container vibrations, also called unjamming, has been extensively studied quantitatively in the past.
Vibrations have also been applied during avalanches to study inhibiting or supporting effects on discharge rates
\cite{Takahashi1968,Suzuki1968,Lindemann2000,Chen2006,Kumar2020}.

The amount of grains discharged between two clogs, the so-called avalanche size $S$, is one of the key figures of merit in silo discharge.
Avalanche sizes are statistically distributed.
Their mean size $\langle S\rangle$ increases with increasing outlet width, more specifically with increasing ratio $\rho$ of orifice width to particle size.
Empirically, a power law has been proposed for the relation $\langle S\rangle(\rho)$ \cite{Zuriguel2005}.
A crude estimate of the orifice width necessary for free uninterrupted flow of hard spheres is five times the particle diameter
\cite{Zuriguel2005,Mankoc2009}.
On the other hand, an exponential dependence was derived theoretically by Thomas and Durian \cite{Thomas2015} on the basis of a microscopic model.
In their description, there is no critical orifice size that sharply
separates the free-flowing from the clogging regime \cite{Thomas2015}. However, the avalanche sizes become very large when $\rho$ exceeds a value of
approximately 5. For typical avalanche sizes found in experiments, the two models are practically indistinguishable.
The Beverloo equation describes the discharge rate both during avalanches and in the free flow regime.

Most experiments and numerical simulations have been performed with hard monodisperse spherical grains in the past. Few experiments
of non-spherical shapes \cite{Zuriguel2005,Borzsonyi2016,Ashour2017,Szabo2018} demonstrated that many features of the discharge can be
compared to those of spheres when an effective particle radius is introduced. Examples of numerical simulations of silo discharge
of non-spherical particles with multisphere-DEM include rods \cite{Torok2017} and mixtures of spherical and
rodlike grains \cite{Lattanzi2019}.

Many features of silo discharge are not only qualitatively but even quantitatively similar in two-dimensional (2D) and three-dimensional geometries.
The 2D container geometry offers the advantage that internal structures and dynamic processes can be directly observed with non-invasive
optical techniques.
Interesting features are the spontaneous formation of blocking arches \cite{Tang2011,Zuriguel2014_1}, the preceding kinetics
\cite{Rubio-largo2015}, and the identification of force networks in the blocking structures \cite{Hidalgo2013,Vivanco2012}.
The detailed structure and stability of clogs of hard grains has been investigated  \cite{Drescher1995I,Drescher1995II,Manna2000,Pugnaloni2001,To2001,To2002,Zuriguel2003,Pugnaloni2004,Lozano2012,Garcimartin2013}.
For 3D systems, a few results on clogging structures were obtained using X-ray imaging \cite{Borzsonyi2016,Torok2017}.

It was also shown that non-adhesive colloidal particles in suspension flow across constrictions in a way that is very similar to the
behavior of  dry non-cohesive granular materials \cite{Marin2018,Souzy2020}.

Silo discharge of hard grains has been investigated in numerous studies in the past, yet
in agriculture, pharmacy and many technological processes, one frequently encounters soft particles. Nevertheless,
soft granular material has been investigated only scarcely so far \cite{Hong2017,Ashour2017b,Stannarius2019,Stannarius2019b}.
In earlier experiments, gas bubbles in a liquid \cite{Bertho2006} and liquid droplets in an oily emulsion \cite{Hong2017b,Hong2017}
were investigated. Hydrogel spheres (HGS) served as elastic solids \cite{Hong2017,Ashour2017b,Stannarius2019,
Stannarius2019b}. These HGS are incompressible and moderately deformable, with elastic moduli of the order of 10 kPa to 100 kPa.
As a descriptive measure of this softness, one can regard the amount of deformation
induced by the pressure at the silo bottom. In a silo of about 1~m fill height, as used in our experiments, the particles
are compacted by up to about 20\% of their diameter (compensated by a transverse expansion because of volume conservation).

Several peculiarities distinguish silo discharge of this soft material from that of hard grains. First, it is a striking new feature
that these low-friction soft grains do hardly clog, even when the orifice size is only slightly larger than two particle diameters
\cite{Hong2017,Ashour2017b}.
Only below that size, the system forms clogged states. Second, the very low friction coefficient of these HGS leads to a nearly
hydrostatic pressure characteristics, at least at practical fill heights up to about 100 particle diameters \cite{Ashour2017b}.
This pressure in the quasi-2D hopper is linearly related to the fill height.
In contrast to rigid grains, there is also a clear fill-height dependence of the discharge rate and even of the discharge characteristics.
For each orifice size below two particle diameters, one finds a characteristic fill height at which the outflow stops permanently \cite{Ashour2017b}.
An aspect that has been disregarded in the earlier studies is the occurrence of transient clogs
in these systems. These are blocking structures that dissolve spontaneously after some time. Such structures have been found earlier in other
contexts, viz. oil droplets in emulsions \cite{Hong2017b}, living matter like animals passing a gate \cite{Zuriguel2014} or pedestrians \cite{Helbing1995,Helbing2000b,Helbing2005}.

In the present study, we analyze the flow of soft, low-friction grains through small orifices and the spontaneous formation and
dissolution of non-permanent congestions of the outlet.

\section{Experimental setup}

\label{sec:Setup}

\begin{figure}[htbp]
    \centering
    \includegraphics[width=0.4\textwidth]{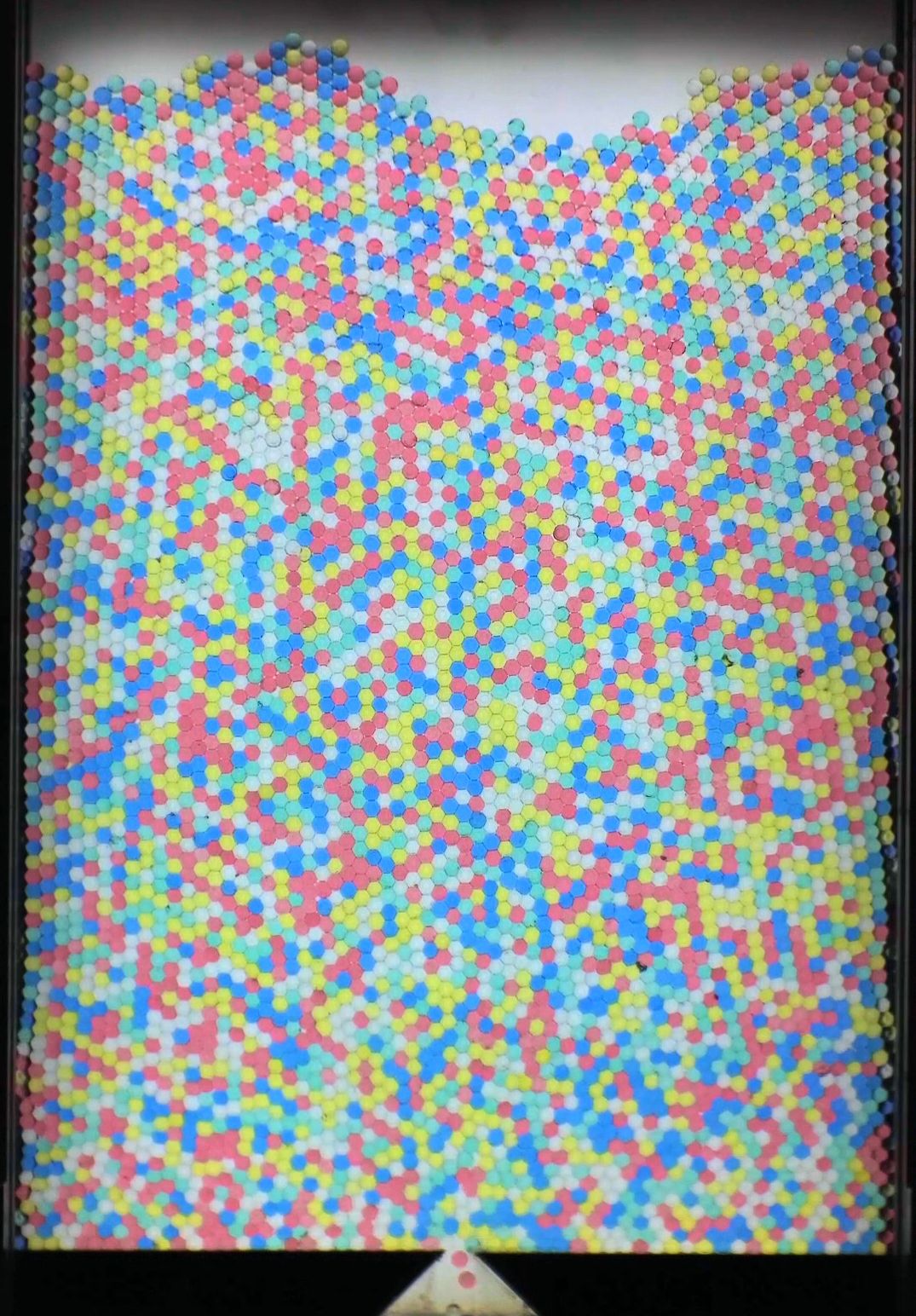}\\
    \smallskip
     \centerline{\includegraphics[width=0.22\textwidth]{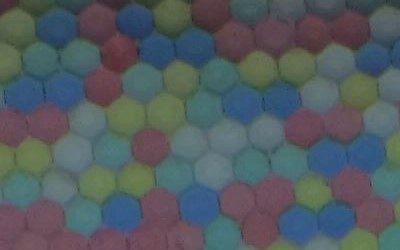}
      \includegraphics[width=0.22\textwidth]{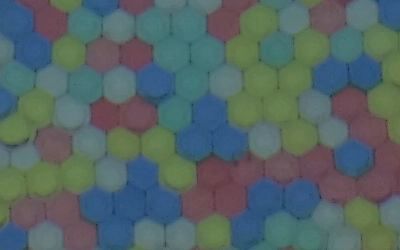}}
\caption{Image of the 2D silo filled with 6.5 mm HGS particles, snapshot during discharge (33 sec after initiation).
Grains with different colors are equivalent, colors are only used to trace individual grains to monitor flow processes. The bottom images show expanded details
of regions 1~cm below the upper layer (left) and 1~cm above the bottom (right). The orifice width is 12 mm.}
    \label{fig:setup}
\end{figure}

The setup consists of a flat box of 80 cm height and 40 cm width, slightly thicker than the particles' diameter, with an adjustable orifice in the center of the bottom plate \cite{Ashour2017b}.
A front view is shown in Fig.~\ref{fig:setup}. In the images, the side edges are hidden by 3 cm wide aluminum bars carrying the front and rear glass plates,
so that the optically accessible area in the pictures is only 34 cm wide. Two symmetric sliders at the bottom are used to fix the orifice width $W$.
They are tapered towards their ends.
The bin can hold about 9,500 grains with a total weight of $\approx 1.75$ kg.
At the top of the container, there is an additional storage volume that can comprise another $\approx 2.5$~kg of grains.

Hydrogel spheres were acquired from a commercial supplier ({\em Happy Store, Nanjing})
in dry form. They were soaked before the experiments for at least 24 h in a NaCl solution. The final size of the swollen HGS depends upon the salt
concentration which was chosen such that we obtained uniformly-sized HGS with 6.5~mm diameter, which varied by approximately 3~\%.
The mass of a single HGS is 0.185~g. The friction coefficient is very low,
of the order of 0.01 or lower. The elastic modulus is approximately 50 kPa (they are slightly softer in the outer shell than in their cores).
We determined the elastic moduli from diameters of Hertzian contacts under given weights. A reasonable measure of the softness of the particles
is the ratio of the pressure at the bottom of the container (of the order of a few kPa for fill heights of up to 1 m) and the elastic modulus.
In our setup, this ratio can reach a value of 0.1.

The setup is observed with a commercial video camera (XiaoYi 4K+ Action Camera), and videos are taken with a frame rate of 60 fps and, if not stated otherwise, with
a spatial resolution of 0.327 mm/pixel.
Below the orifice, we collect the discharged particles in a tray mounted on a balance. 
The HGS are taken from the storage bath, placed on tissue to remove excess water and then filled into the silo from above, while the orifice
is closed.
After filling, the orifice is opened and we record in parallel the weight of the discharged material
and the video of the silo front side. Video and mass measurements can be synchronized better than 0.5 s.

In order to demonstrate the differences of the behavior of the soft, low friction grains in comparison to hard frictional particles of a
similar size and density, we have performed some experiments with hard plastic ammunition (Airsoft bullets, ASB).
These have a similar density as the HGS and a comparable diameter of 6~mm. ASB have a friction coefficient of approximately 0.3. Their deformability
can be neglected in our experiments.

\begin{table}
\begin{tabular}{|l|l|l|l|}
  \hline
  Material & Orifice& $\rho$ & Character of discharge\\
  \hline
  ASB & 34 mm & 5.7  & free flow \\
  HGS & 18 mm & 2.77 & free flow  \\
  HGS & 15 mm & 2.30 & free flow\\
  HGS & 12 mm & 1.85 & fluctuating flow, $h_c\approx 4$ cm\\
  HGS & 11 mm & 1.69 & intermittent flow, $h_c\approx 11$ cm\\
  HGS & 10 mm & 1.55 & intermittent flow, $h_c\approx 23$ cm\\
  \hline
\end{tabular}
\caption{Materials, orifice sizes, and the character of discharge. $\rho$ is the ratio of orifice width
(gap between upper edges of the two sliders,
see Fig.~\ref{fig:setup}) and particle diameters. In cases where the system clogs permanently,
$h_c$ is the average remaining fill height.\label{tab:1}}

\end{table}

Table \ref{tab:1} lists the combinations of orifice sizes and materials in this study, and the
character of the discharge. Free flow means an uninterrupted discharge where the flow rate is either constant or decreases with decreasing
fill heights. Fluctuating flow is identified by randomly varying flow rates, superimposed on the general trend to lowering
rates with decreasing fill heights. Intermittent flow is characterized by phases where flow is completely interrupted. The distinction
between intermittent and fluctuating flow cannot be defined sharply. There is no clear criterion, since the fluctuations can become large
enough to practically stop the flow. Empirically, we may set some threshold time where the observer by eye does not perceive motion of the grains
blocking the orifice. These details will be clarified in the following sections.

\section{Experimental results}

\subsection{Packing structures}
The ASB are packed in domains of a perfect hexagonal lattice with few defects and dislocations forming the domain borders. The effective 2D packing fraction $\phi_{2D}$
(in the container midplane) is close to the maximum packing density  $\phi_{2}=\pi/\sqrt{12}\approx 0.9069$ for identical disks in a hexagonal 2D lattice.
During the discharge, the packing fraction lowers noticeably in the flowing region, in particular near its edges and directly above the orifice, owing to Reynolds dilatancy.

The situation for the HGS particles is quite different. They also form a hexagonal lattice, but practically without dislocations or defects.
However, there are long-range distortions of the lattice (see Fig.~\ref{fig:setup}).
Since the HGS can be substantially deformed, the 2D packing fraction $\phi_2$ is, although still applicable, not the most reasonable presentation. While we determine
the local packing densities by counting the spheres per area, we present the results in terms of the 3D packing fraction $\phi_{3D}$.
The densest packing of uniform spheres with diameter $d$ in a cell with thickness $d$ yields
$\phi_{3}=\pi/\sqrt{27}=\frac{2}{3} \phi_{2} \approx 0.6046$.

It is evident that the packing structure of the soft HGS is denser at the bottom than at the top (see Fig.~\ref{fig:setup}, bottom). Moreover, the packing fraction depends in a complex fashion on the history of the ensemble.
After filling the hopper, we counted a 2D packing density of approximately 3.07 spheres/cm$^2$ in an upper part, approximately 45 cm above the orifice (still
approximately 30 cm below the top of the granular bed).
This corresponds to a space filling of $\phi_{3D}=0.683\approx 1.13~\phi_{3}$. The spheres are flattened and squeezed out of the central plane so that the distance between neighbors is smaller than the original sphere diameter. At the bottom of the container, approximately 50~cm below the top of the granular bed, we found on average 3.55 spheres/cm$^2$ ($\phi_{3D}=0.792\approx 1.31~\phi_3$), i.~e. more material of the HGS was squeezed out of the midplane.
The pressure in HGS-filled cells has been measured earlier by Ashour \cite{Ashour2017b}. It was found that the static pressure at the bottom of the silo
grows nearly linearly with height, at a rate of about 7 kPa/m.

As soon as the outflow starts, the packing fraction gradually decreases, starting from the region near the orifice. During the phases of continuous
flow in a hopper with 10 mm orifice, the average packing fraction near the bottom ($\approx 5$~cm above the orifice) dropped by more than 15~\% to about 3.0 spheres/cm$^2$ ($\phi_{3D}=0.665 \approx 1.1~\phi_{3}$) and near the top layer by approximately 5 \% to about 2.93 spheres/cm$^2$ ($\phi_{3D}=0.653\approx 1.08~\phi_{3}$).
We will show in more detail in the next section how the packing fraction relaxes during the discharge.
The packing fraction fluctuates during the discharge by a few percent until the outlet gets blocked. After a longer (several seconds) stagnation of
the outflow, the average packing fraction near the bottom is increased again to roughly 0.756 (approximately $1.25~\phi_3$), and it drops again as soon as the discharge continues.


The focus of the following experiments will be the fluctuations of the outflow that are related to the elastic and frictional characteristics
of the HGS. First, we will concentrate on the flow field inside the container and the reorganization of the packing structures, before
we analyze the dynamics at the orifice. Note that in both ASB and HGS experiments, we used monodisperse ensembles. As stated above, this
leads to more or less perfect lattice structures with dislocations in the container. This influences the flow field particularly in the hard
particle experiment. Polydisperse mixtures may differ in some features.

\subsection{Flow inside the silo}

\begin{figure}[htbp]
     \centerline{
    \includegraphics[width=0.25\textwidth]{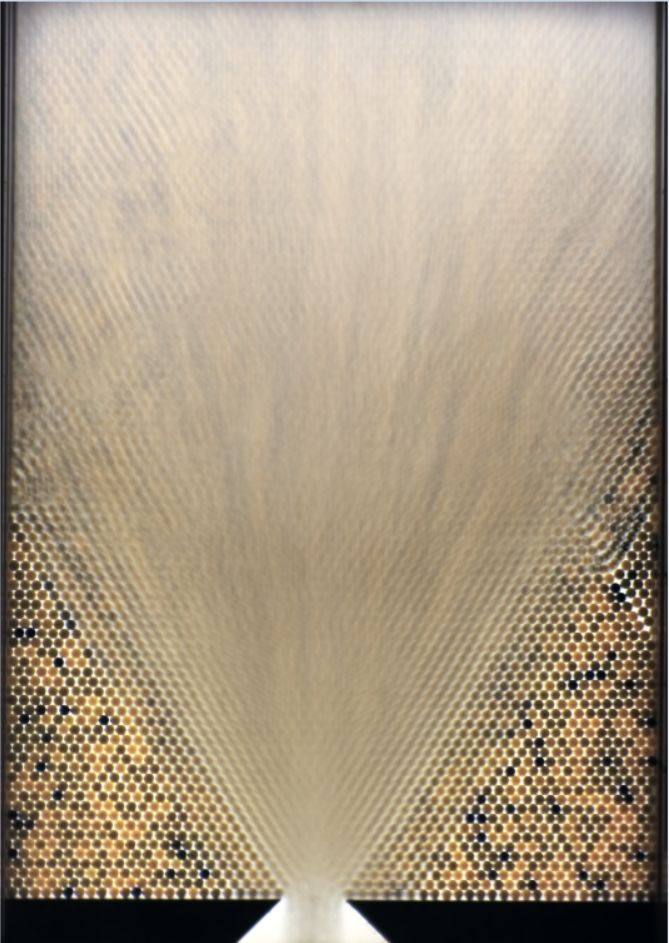}
    \includegraphics[width=0.25\textwidth]{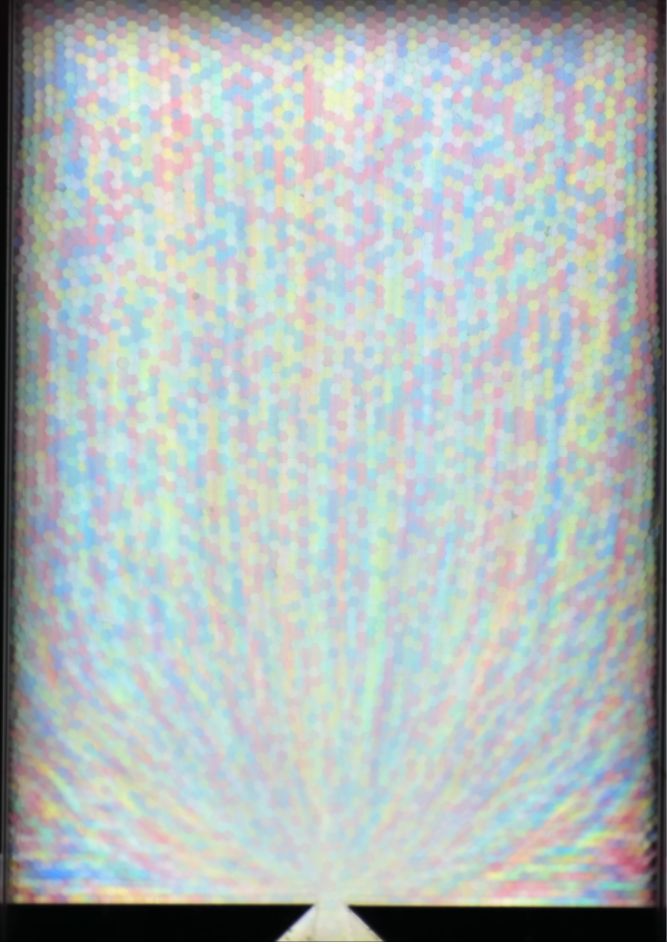}
}
 a)\hspace{0.23\textwidth} b)\hfill
\caption{Superimposed 1000 images ($ \approx 16$~sec) of the 2D silo filled with (a) 6 mm ASB and (b) 6.5 mm HGS particles during the discharge.
The orifice size is 34~mm for the ASB, 18~mm for the HGS. }
    \label{fig:flow}
\end{figure}

The structure of the flow field \cite{SI} can be directly visualized by averaging subsequent images of the recorded videos. Figure \ref{fig:flow}a
shows averages of 1000 frames of the front view of ASB. The image is blurred in flowing regions. The most characteristic feature is that there
is a pronounced flow along the trigonal lattice planes, indicated by ray-like lines, most prominent at the lateral edges of the flowing zone.
A comparable average is shown in Fig. \ref{fig:flow}b for the soft HGS.
Qualitatively, the flow profiles inside the container do not depend significantly upon the orifice sizes, but strongly on the material
properties.

In the ASB-filled silo, two stagnant zones are identified by their stationary hexagonal lattice. At the right hand side, one sees a
dislocation in the lattice structure. In the HGS, the flow is slower in the corners, but there is no stagnant region at all. This finding
is in perfect agreement with observations by X-ray tomography in 3D containers \cite{Stannarius2019,Stannarius2019b}.

\begin{figure}[htbp]

  a)  \includegraphics[width=0.46\textwidth]{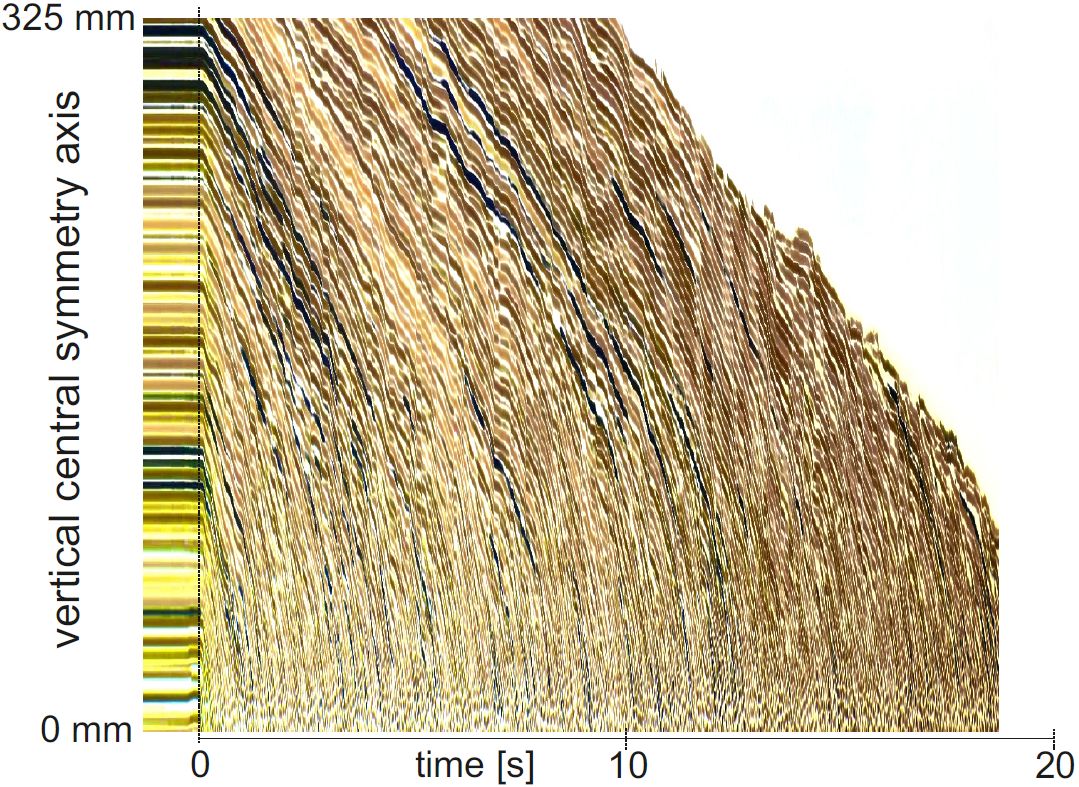}\\ \medskip
  b)  \includegraphics[width=0.44\textwidth]{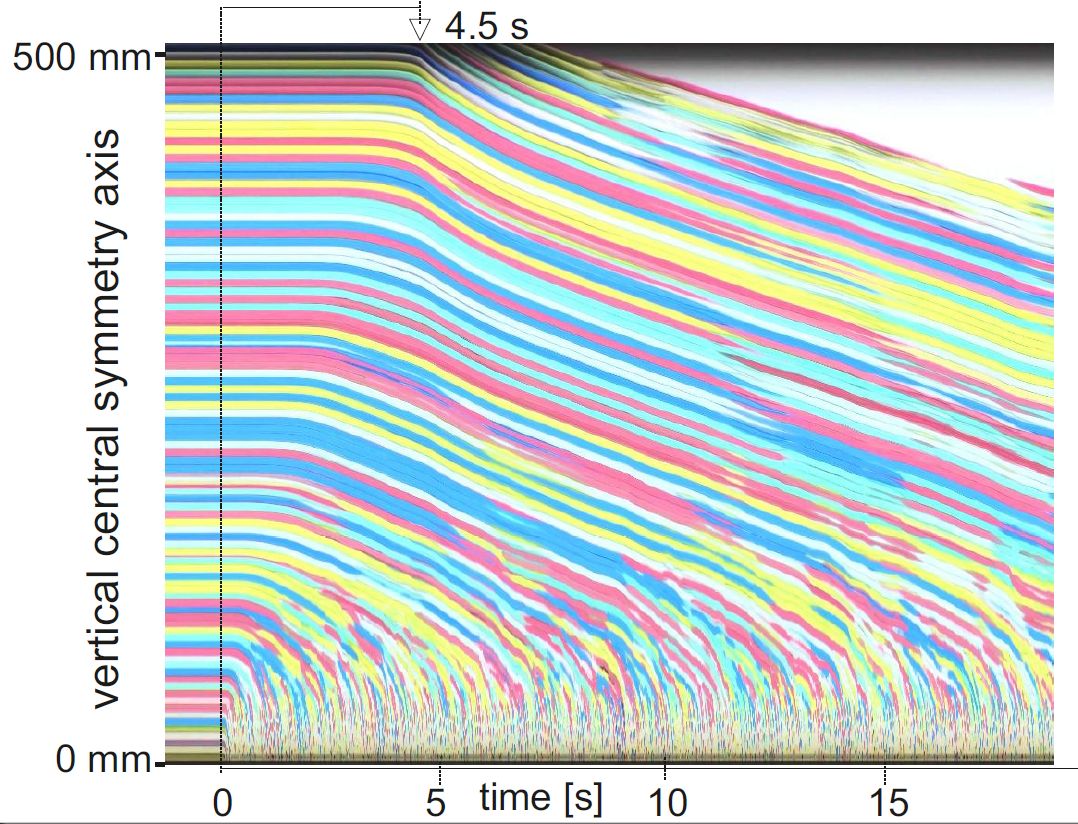}

\caption{Space-time plots of a vertical cut at the central symmetry axis of the silo, (a) ASB, 34~mm orifice width, (b) HGS, 18~mm orifice width.
In the hard particle system, the flow reaches the upper edge of the selected region within approximately 100 ms. In the soft material, it takes
4.5 seconds after the orifice was opened for the flow to reach the upper region shown in the plot. This is even more pronounced at lower orifice
widths because of the lower flow rates.}
    \label{fig:dilat}
\end{figure}
The soft, low-friction grains remarkably differ from hard particles during silo discharge in another aspect: After the start of the discharge, the local packing
fractions change considerably. In the silo filled with hard grains, this quantity changes only marginally, primarily in the sheared regions
near the edges of the flowing zone and in the direct vicinity of the outlet. As a consequence, the flow starting at the outlet causes motion
practically instantly in all layers.
This is visualized in Fig.~\ref{fig:dilat}a. The image shows the space-time plot of a vertical cut along the central vertical symmetry axis of the
silo, above the outlet. After the orifice is opened at time zero (vertical line), the positions of all grains stacked along this cut start
to move down almost immediately. In different experiments, we found delays no longer than 0.1 and 0.2 seconds. The behavior of the soft, low-friction
grains (Fig. \ref{fig:dilat}b) is in sharp contrast. Initially, flow sets in only in the vicinity of the outlet, while the positions of the upper
grains remain unchanged. The material expands at the bottom first, lowering the effective 2D packing fraction as stated above.
This is obviously a consequence of the viscoelastic properties of the HGS. Only after
a substantial amount of material has flown out, in this case roughly 150 g or nearly 1000 spheres, the granular bed has sufficiently diluted and the
flowing region has reached the height of 50 cm, the upper edge of the image.
From then on, the material flows with roughly uniform velocity in the central axis. The downward flow
accelerates only within the final 10 cm above the orifice, where the flow velocity is slower at the sides (cf. Fig.~\ref{fig:flow}b).

\subsection{Outflow velocity and fluctuations in the outflow rate}

\begin{figure*}[htbp]
    \centerline{
     a)\includegraphics[width=0.33\textwidth]{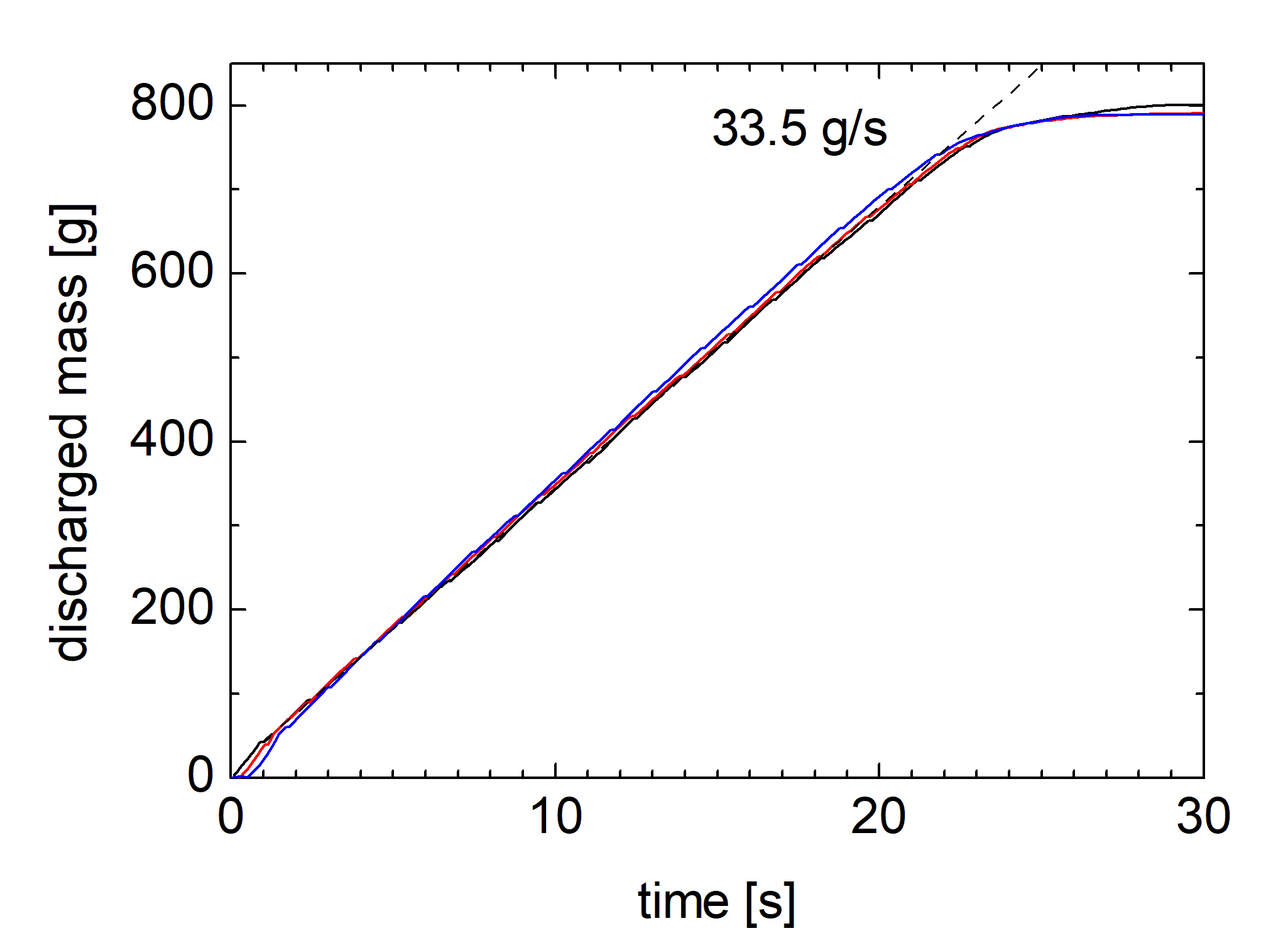} \hfill
     b)\includegraphics[width=0.33\textwidth]{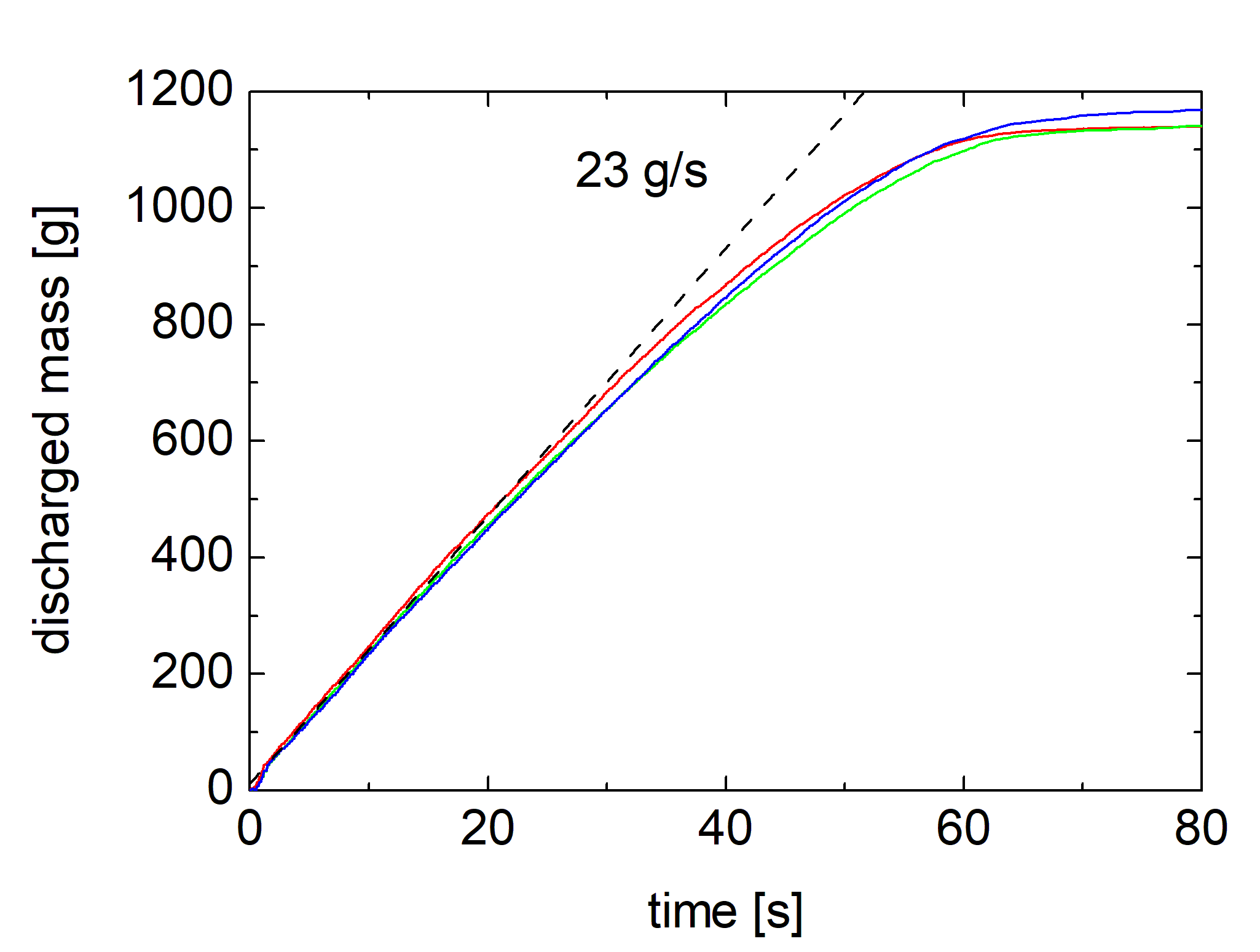}\hfill
     c)\includegraphics[width=0.325\textwidth]{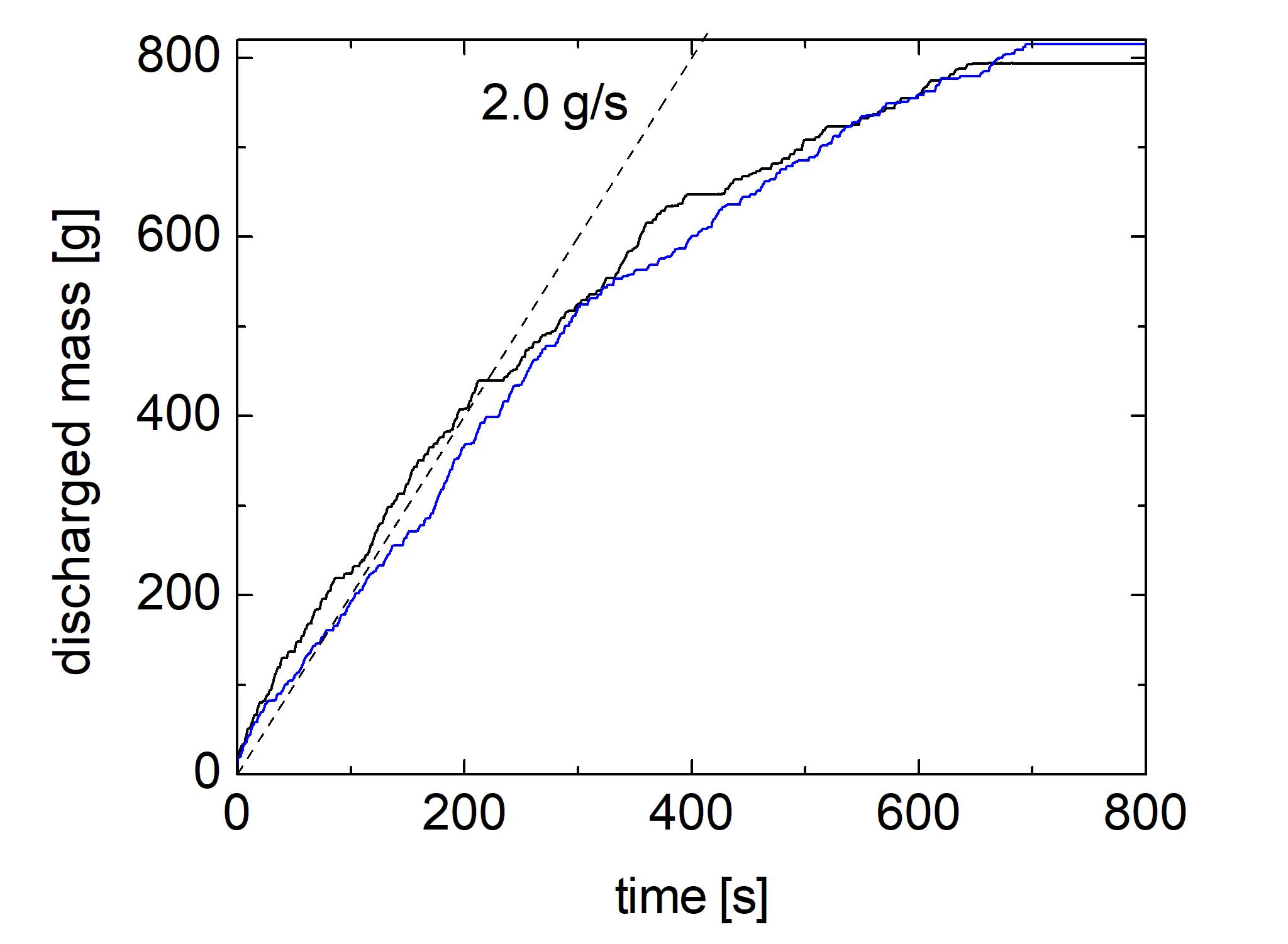}
}

\caption{Discharged mass (a) in the ASB-filled silo with 34 mm orifice width, $\rho=5.7$, (b) in the HGS-filled silo with 18 mm orifice width,
$\rho=2.77$, and (c) in the HGS-filled silo with 10 mm orifice width, $\rho=1.55$. Apart from the very different
time scales caused by the different orifice widths, it is evident that the flow of the hard ASB grains is practically constant until the very
bottom of the silo is reached. In the 18~mm orifice, HGS-filled silo, the discharge slows down with sinking fill level. This is even more pronounced
 in the 10~mm orifice silo where strong flow rate fluctuations occur even when the silo is still well-filled. Different colors distinguish
 individual runs of the experiment.}
    \label{fig:1}
\end{figure*}

The most significant difference between the flow of hard and soft grains is the temporal variation of the outflow rate.
Figure \ref{fig:1}a shows the outflow of rigid ASB from the same silo, where we have set the orifice size to 34 mm (orifice to particle size ratio $\rho=5.7$).
It represents the standard behavior of hard particles. We find a continuous
outflow until the silo is nearly emptied. The outflow slows down only when the silo is almost empty and the stagnant
zones erode from above until the static angle of repose is reached. Finally, some
material remains in the bottom corners. At smaller orifice sizes, the same features are found, except that the discharge will stop completely
when a permanent clog blocked further outflow.

\begin{figure}[htbp]
    \centerline{
     \includegraphics[width=0.45\textwidth]{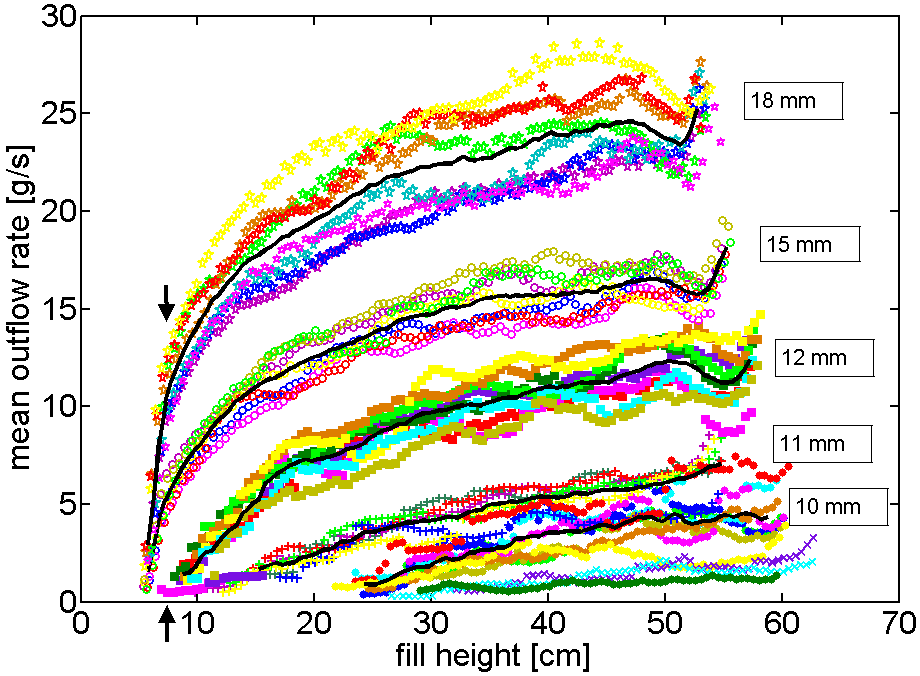}
}
\caption{Mean outflow rates of HGS for different orifice sizes as a function of the instantaneous fill level. All rates are averaged
over 5 cm height difference. The solid curves represent the averages of curves belonging to one set of experiments at fixed orifice
sizes. Arrows indicate where the tip of the v-shaped surface of the granular bed reaches the outlet.}
    \label{fig:rate}
\end{figure}

In the HGS-filled silo with large enough orifice sizes, the behaviour is similar to that of the hard grains, but there is a slight pressure dependence
of the outflow rate. This becomes more evident with decreasing orifice size. When the orifice is approximately 3 particle diameters wide (Fig.~\ref{fig:1}b), there is still continuous outflow of the HGS without clogging.
However, one observes a clear pressure dependence (dependence upon the instantaneous fill height)
of the discharge rate, even when the silo is still half-filled (cf. Fig.~\ref{fig:rate}). For comparison, hard ASB will permanently clog at this relative orifice size with
mean avalanche sizes of only 65 grains ($\approx 8$~g) \cite{Ashour2017}.

When the orifice size is further decreased, the dependence of the
mean outflow rate on fill height gets more pronounced. In addition, the outflow rate starts to fluctuate. This is evident in Fig.~\ref{fig:1}c where
we show the discharge through a 10 mm orifice ($\rho \approx 1.55$). Not only is the initial flow rate reduced respective to the 18 mm orifice
by one order of magnitude, but the discharge curve also shows clear plateaus where the outflow stops for several seconds.

\begin{figure*}[htbp]
    \centerline{
     a)\includegraphics[width=0.33\textwidth]{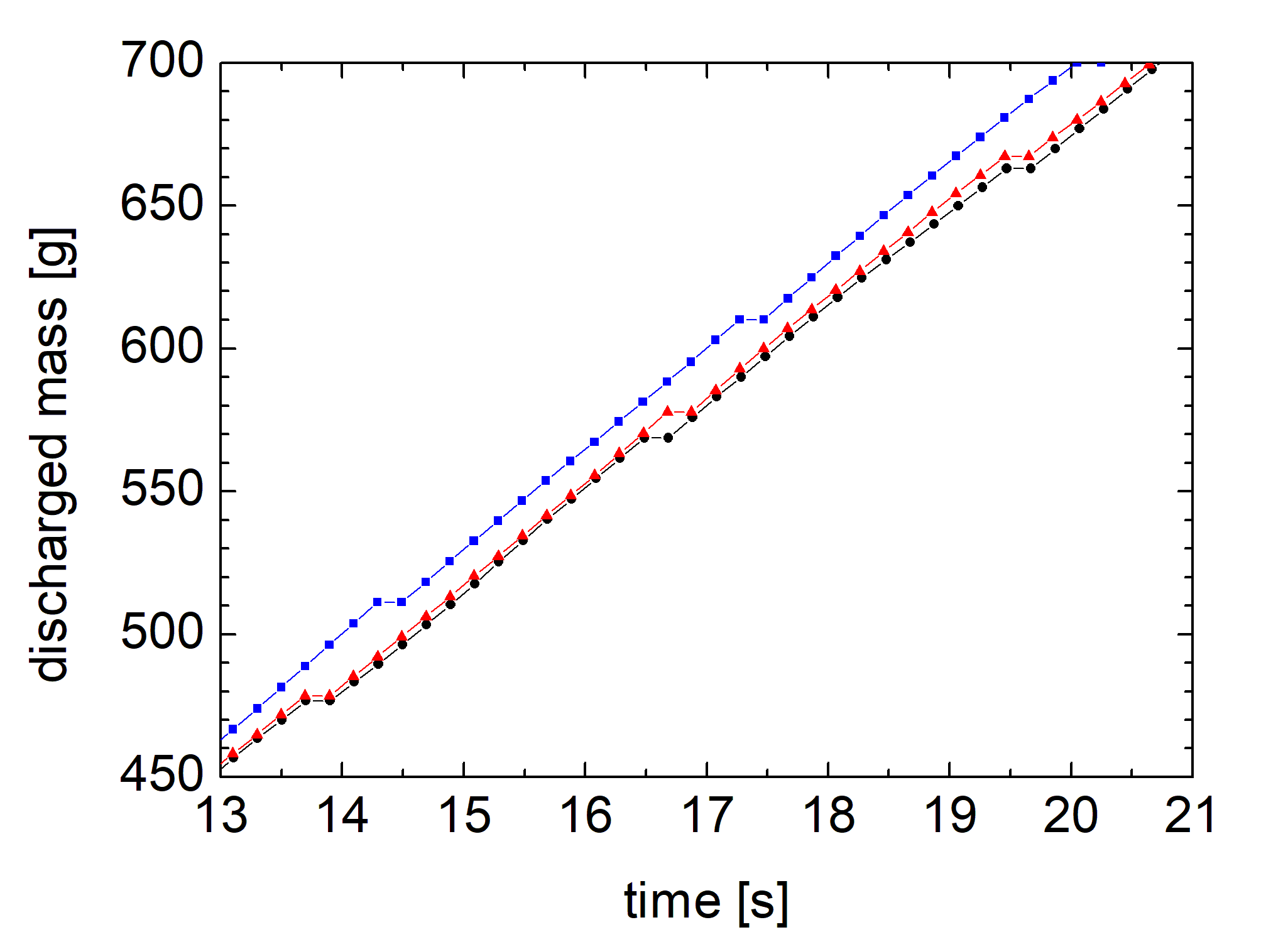} \hfill
     b)\includegraphics[width=0.33\textwidth]{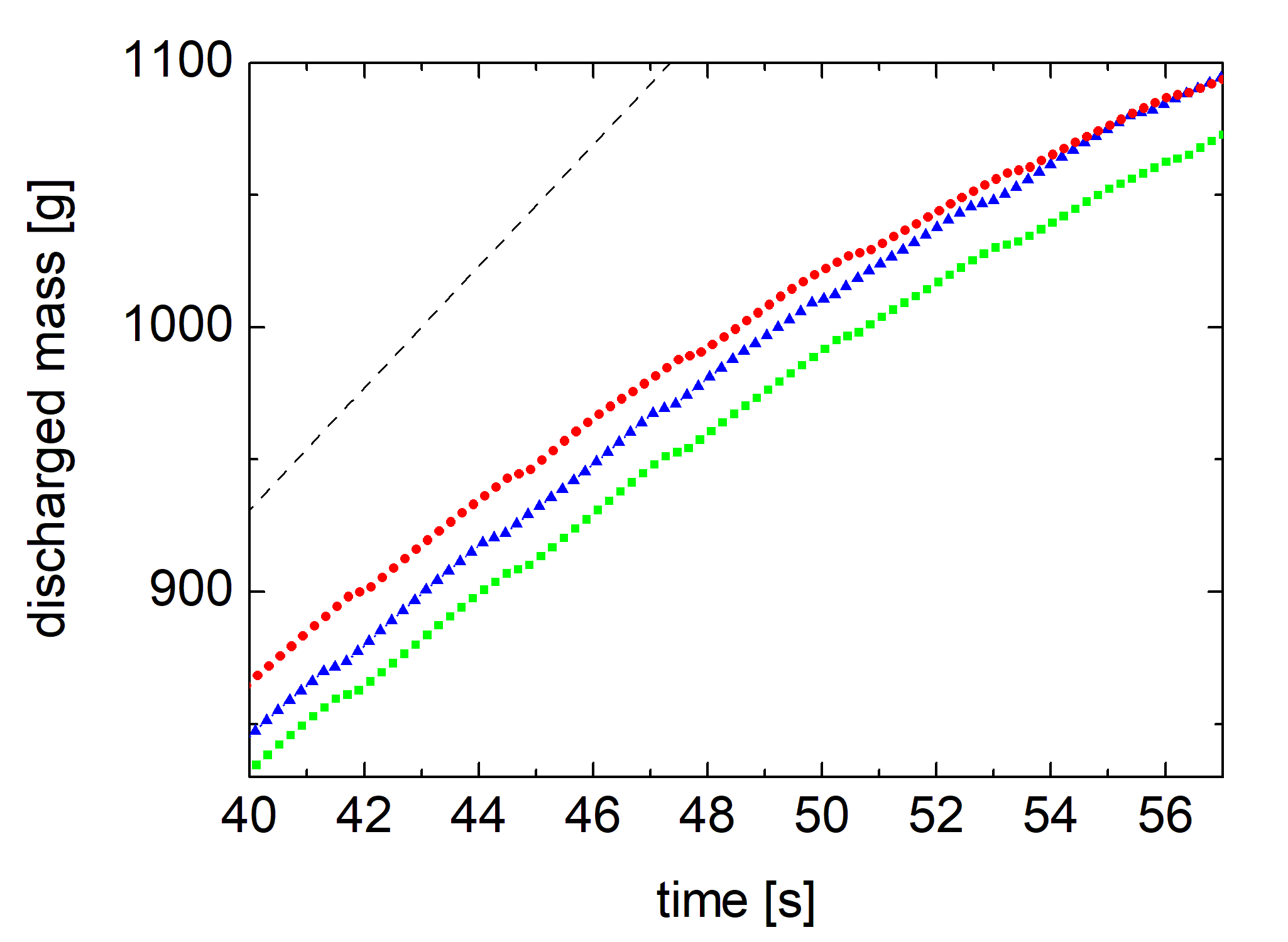}\hfill
     c)\includegraphics[width=0.333\textwidth]{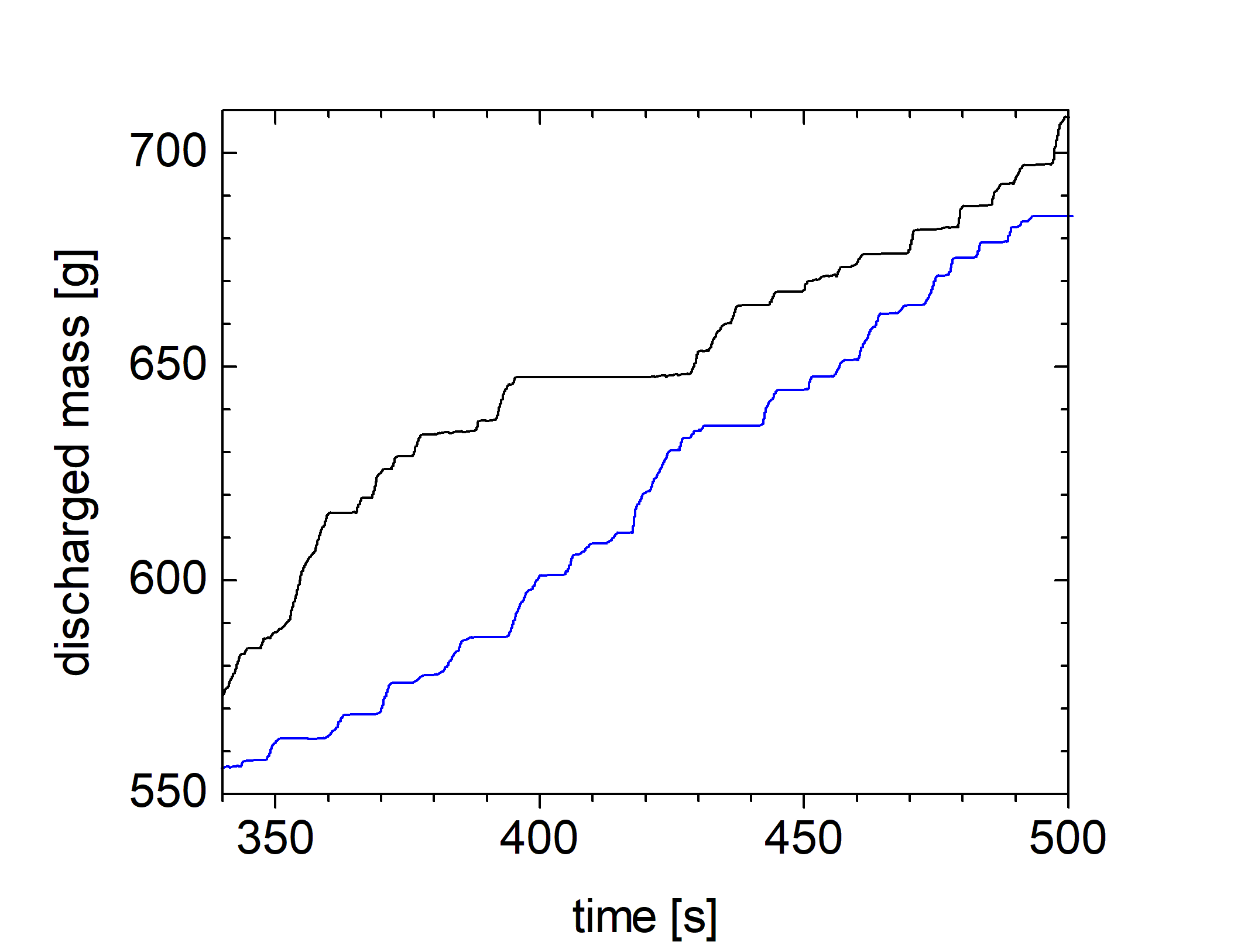}
}
\caption{Zoomed-in details of the three graphs shown in Fig. \ref{fig:1}(a-c). All graphs show periods ending approximately 100 g before
the discharge stops (empty silos in (a) and (b), permanent clog in (c)). The small step-like undulations in the graphs (a,b)
are a beating artefact of the sampling frequency of the balance (4.66 Hz) and the poll frequency of the computer (10 Hz).    \label{fig:2}}
\end{figure*}

\begin{figure*}[htbp]
    \centerline{
     a)\includegraphics[width=0.20\textwidth,height=0.45\textwidth]{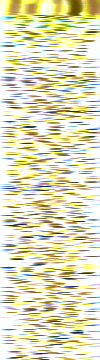}\hfill
     b)\includegraphics[width=0.20\textwidth,height=0.45\textwidth]{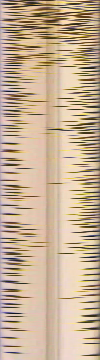}\hfill
     c)\includegraphics[width=0.14 \textwidth,height=0.45\textwidth]{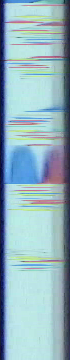}\hfill
     d)\includegraphics[width=0.14 \textwidth,height=0.45\textwidth]{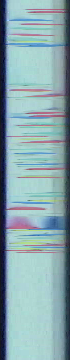}\hfill
     e) \includegraphics[width=0.14\textwidth ]{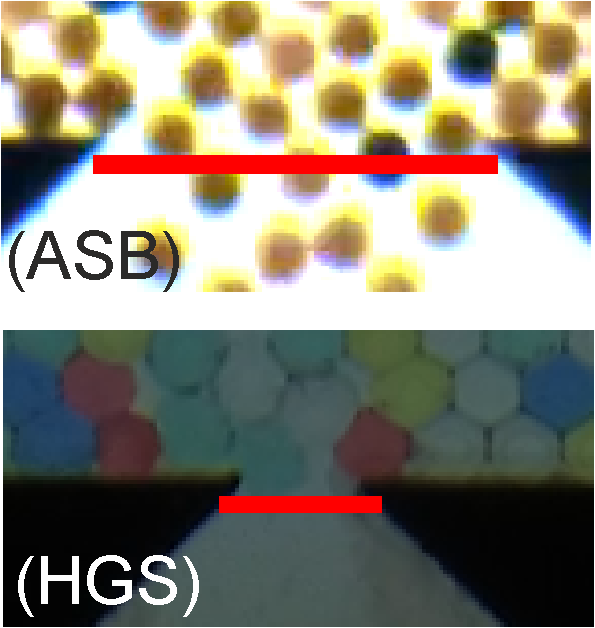}}

\caption{Space-time plots of cross sections directly below the outlet, time runs from top to bottom, each plot covers a time interval of 6 s.
The discharges started at time $t_0$. (a) ASB, $\rho=5.7$, ($t_0-0.3$ s) to ($t_0+5.7$ s), (b) ASB $\rho=5.7$, ($t_0+23$ s) to ($t_0+29$ s),
(c) HGS at $\rho = 1.55$, $t_0$ to ($t_0+6$ s), (d) HGS at $\rho = 1.55$, ($t_0+480$ s) to ($t_0+486$ s). The cuts (a,b)
are 36 mm wide, those in (c,d) are 12.5 mm wide. e) shows the positions of the cross sections at the orifice.}
    \label{fig:3}
\end{figure*}

Zoomed-in details for the same data as in Fig.~\ref{fig:1} are shown in Fig.~\ref{fig:2}. The discharge rate of the hard grains
(Fig.~\ref{fig:2}a) is linear within the experimental resolution of our setup.
Small steps seen in the curves are artifacts of the measurement technique, resulting from the time resolution of the balance:
The balance updates its values every 0.214 s, while the computer samples these data every 0.1 s, thus each balance datum is regularly
read out twice, but three times every 1.5 s.

The 18 mm HGS curve (Fig.~\ref{fig:2}b) is not straight but still smooth except for the readout artifacts.
In clear contrast, the 10 mm HGS plot (Fig.~\ref{fig:2}c) shows strong fluctuations and steps that are no
artifacts (note the different scales of the time axes).

Since the outflow rate is primarily determined by the pressure at the container bottom, it is more instructive to plot the outflow rates depending
on the remaining fill heights instead of time. In Fig.~\ref{fig:rate}, the momentary rates are presented as functions of the instantaneous container
fill level. In order to smooth these graphs, we averaged these data over periods where the fill level dropped by 5 cm. The top edge of the flowing
granular bed is v-shaped, we use the mean height of this edge as the relevant parameter. The arrows in the figure indicate where the tip of the v
has reached the outlet (independent of the orifice widths), and the granular bed splits in two cones left and right of the orifice, from which
particles slide down and pass the gap.

As seen in the figure, all rates are strongly fill-level dependent, and the discharge rates for given orifice sizes vary considerably between
individual runs of the experiment, with no systematic trend.
For narrow orifices in particular, the outflow rates depend sensitively on the preparation of the samples. Even though the HGS surfaces are dried
with tissue before filling them into the silo, slight variations in moisture on the HGS surfaces may affect cohesion of the spheres by capillary
bridges and influence the magnitude of the flow.

Next, we analyze the temporal fluctuations of the outflow. One can directly visualize the differences between hard and soft grains
without having to rely on the limited time resolution of the weight measurement:
We have constructed space-time plots of a horizontal cut just below the orifice in the video frames. Figure \ref{fig:3}
shows exemplary plots for the ASB at 34 mm opening and HGS at 10 mm opening. The ASB pass the orifice rather continuously, practically
independent of the container fill height. The plot (b) of the final phase of the discharge shows that it is not a reduced rate
of particles passing the orifice that flattens the graph in Fig. \ref{fig:1}a near the end. Instead, particles do not pass through the
complete outlet gap anymore. When the stagnant zones erode, grains rolling down the slopes of the remaining piles only pass the
lateral sides of the orifice.

\begin{figure*}[htb]
    \centerline{
     a)\includegraphics[width=0.47\textwidth]{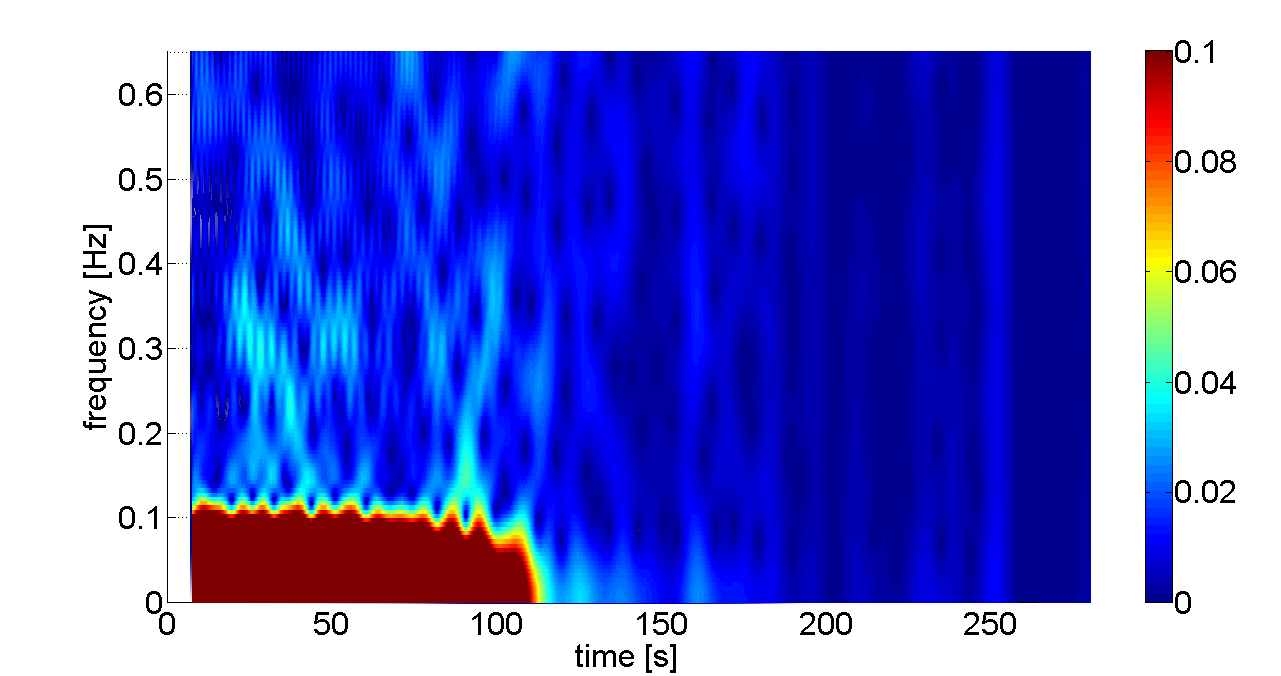}\hfill
     b)\includegraphics[width=0.47\textwidth]{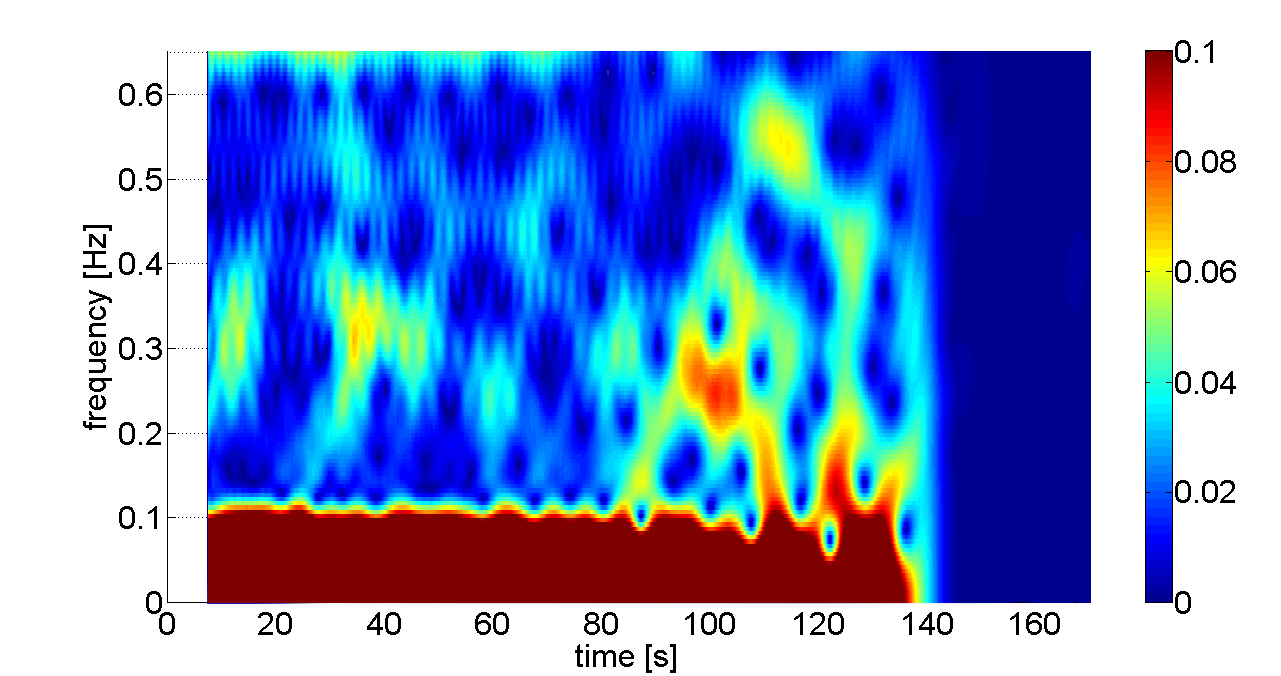}}
    \centerline{
     c)\includegraphics[width=0.47\textwidth]{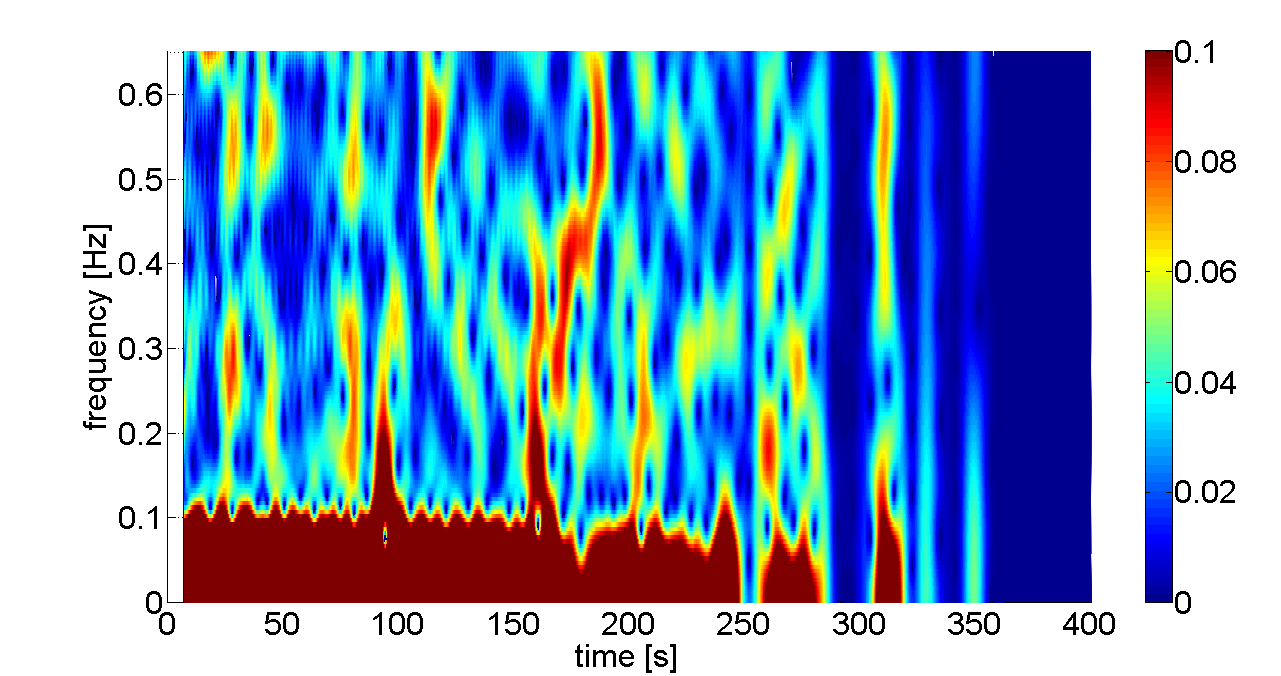}\hfill
     d)\includegraphics[width=0.47\textwidth]{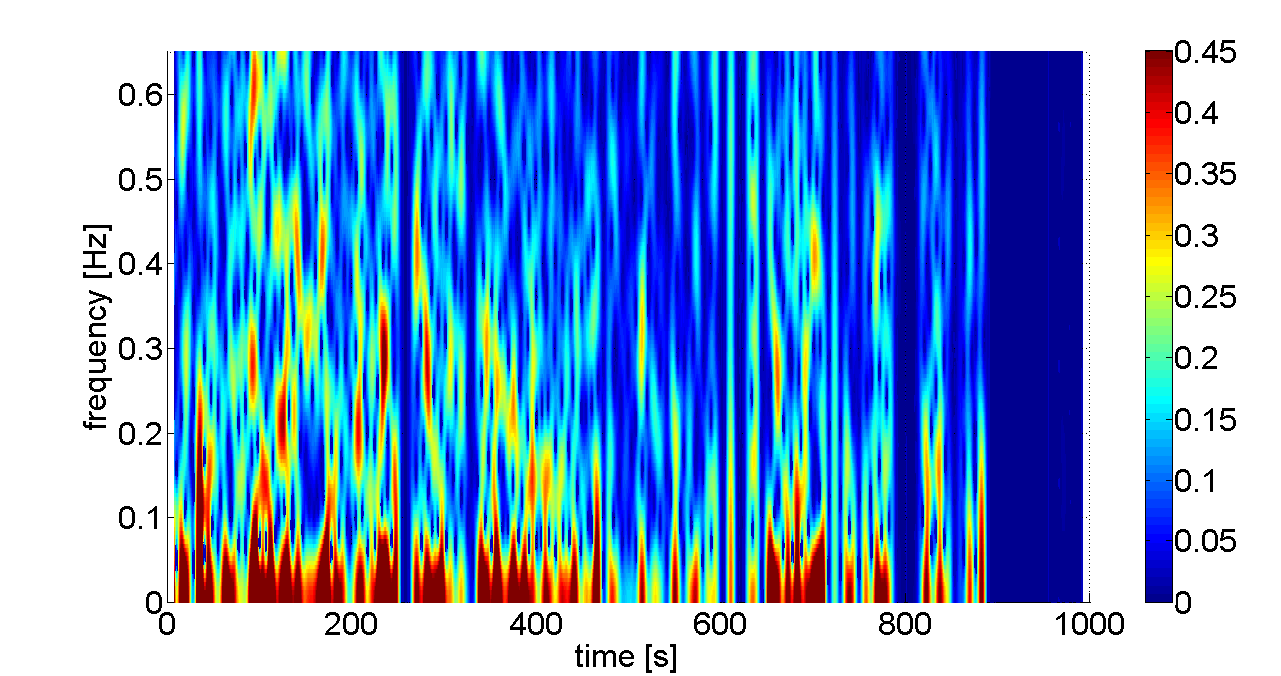}
     }
\caption{Windowed Fourier Transforms (Hamming window width 15 s) of the HGS outflow rates for different orifice sizes, (a) 15 mm, (b) 12 mm, (c) 11 mm, and
(d) 10 mm. The peaks are scaled with the maximum fourier amplitude $q_{\max}$ and clipped at $0.1 q_{\max}$. Note the different color scale in (d).
}
    \label{fig:4}
\end{figure*}

In clear contrast, the passage of soft, low-friction grains through the outlet is fluctuating substantially. At orifice sizes between $\rho \approx 2$
and $2.5$, these fluctuations set in only when the container is nearly empty and pressure is low. At orifice sizes below $\rho=2$,
they are permanently present in our setup. This is seen in the space-time plots of Figure \ref{fig:3}c,d. While there are phases when the
grains leave the opening at a rate comparable to the hard grains, there are clearly distinguished phases where the orifice is blocked
temporarily. These phases where the outflow is clogged completely are distributed in lengths, they can persist for several seconds as
seen in Figs. \ref{fig:2}c and \ref{fig:3}c,d, and they dissolve spontaneously. The mechanism is discussed below.

Most of these transient clogs are formed by arches of 4 particles, whose configurations are not static in time.
From time to time, the 10 mm orifice is blocked by a pair of particles that has already partially passed the
bottleneck (Fig.~\ref{fig:3}c,d). In the space-time plots of cross-sections taken $\approx 2$ mm below the narrowest part of the orifice,
such particles exhibit a bell-shaped signature.
(blue and red particles after about 2.5 seconds in (c), red and blue particles after about 3 s in (d)). Their signature in the plots
evidences that they are continuously squeezed into the gap, and then suddenly drop down, releasing a new avalanche.
The nature of this discontinuous flow makes it difficult to define avalanche sizes for the soft, low-friction grains at narrow orifice sizes, since there
is no clear definition of what can be considered a clog. A similar problem was discussed by Hidalgo \cite{Cruz2018} and Souzy \cite{Souzy2020}
in the context of colloidal systems.

It is important to emphasize that with respect to the internal state of the granular ensemble, the non-permanent congestions that
we observe are not static. Each of the discharge scenarios shown, e. g., in Figs. \ref{fig:1}c, \ref{fig:3}c,d, has to be considered one
single avalanche until the permanent clog has formed. This is explained in detail in the following Section~\ref{sec:NPC}.

The fluctuations in the outflow rate can be characterized quantitatively by windowed Fourier Transforms of the discharge rates that were
extracted from the balance data. We restrict this analysis to the low-frequency range below 0.6 Hz. Thereby, the artifacts from the balance
readout at 2/3 Hz are outside the observed frequency range. We focus on the fluctuations on time scales of several seconds. Figure \ref{fig:4}
shows four different typical frequency spectra. All spectra are normalized with the respective maximum Fourier amplitude
$q_{\max}$ at zero frequency and clipped the low-frequency part of the spectrum for better visibility at $0.1 q_{\max}$.
In none of the spectra, we observed a persistent frequency.
At 18 mm orifice size (not shown),
fluctuations are hardly seen in the spectrum. At 15 mm (Fig.~\ref{fig:4}a), one can observe slight fluctuations with amplitudes of about
4\% of $q_{\max}$. This trend increases significantly for $\rho<2$. At 12 mm orifice width (Fig.~\ref{fig:4}b), the fluctuations
considerably intensify towards the end of the discharge. Fig.~\ref{fig:4}c shows that at 11 mm orifice width, the fluctuations reach
already 10 \%, even at the beginning of the discharge. In Fig.~\ref{fig:4}d, for the smallest orifice size of 10 mm ($\rho=1.55$),
fluctuations reach more than 40 \%, and some interruptions extend over the full width of the Fourier Transform time window,
i.e. some congestions last longer than 15 s.

\subsection{Non-permanent congestions}
 \label{sec:NPC}

We will now analyze in more detail what causes the interruptions and restart of the outflow through small orifices ($\rho<2$),
seen for example in Figs. \ref{fig:2}c, \ref{fig:3}c,d and \ref{fig:4}c,d and in the supplemental video \cite{SI}. The phenomenon of spontaneous non-permanent clogging
has been described earlier for active matter like pedestrians and animals and for thermal systems like colloids \cite{Cruz2018}.
Zuriguel et al. \cite{Zuriguel2014} have coined the term 'clogging transition' for such
scenarios. A characteristic figure of merit of such transitions is the distribution of durations of clogged states, which was found to
follow a power law for sufficiently long clogs. On the other hand, a problem is the clear distinction between delays between
passages of individual particles and short clogs.
The definition of a clogged state in these systems can influence the avalanche statistics substantially \cite{Cruz2018}.
The difference to the present soft granular system is the existence of thermal noise or activity of the involved entities in those systems
which can destroy an existing clog.
\begin{figure}[htpb]
    \centerline{
     \includegraphics[width=0.5\textwidth]{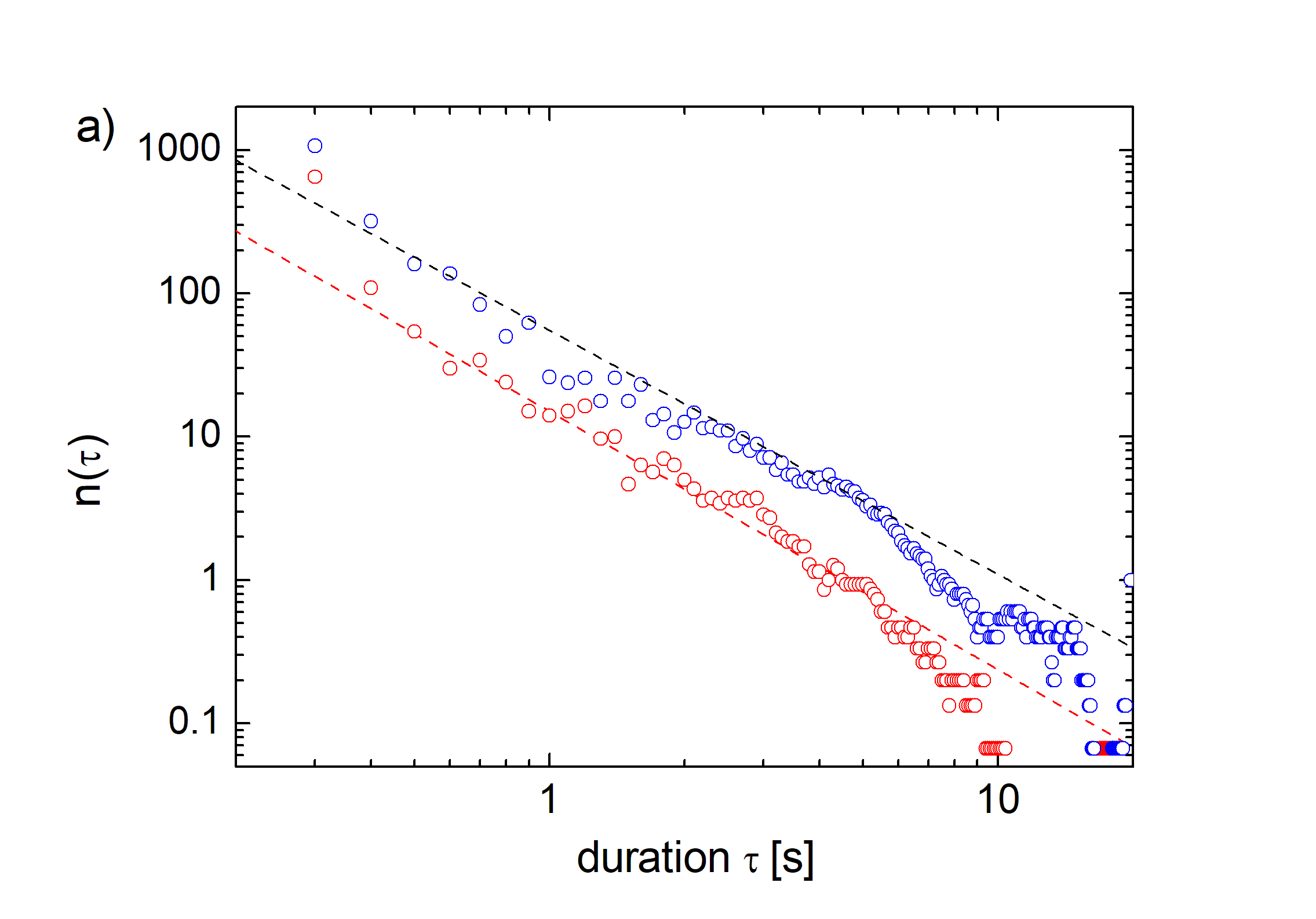}
     }
     \centerline{
     \includegraphics[width=0.5\textwidth]{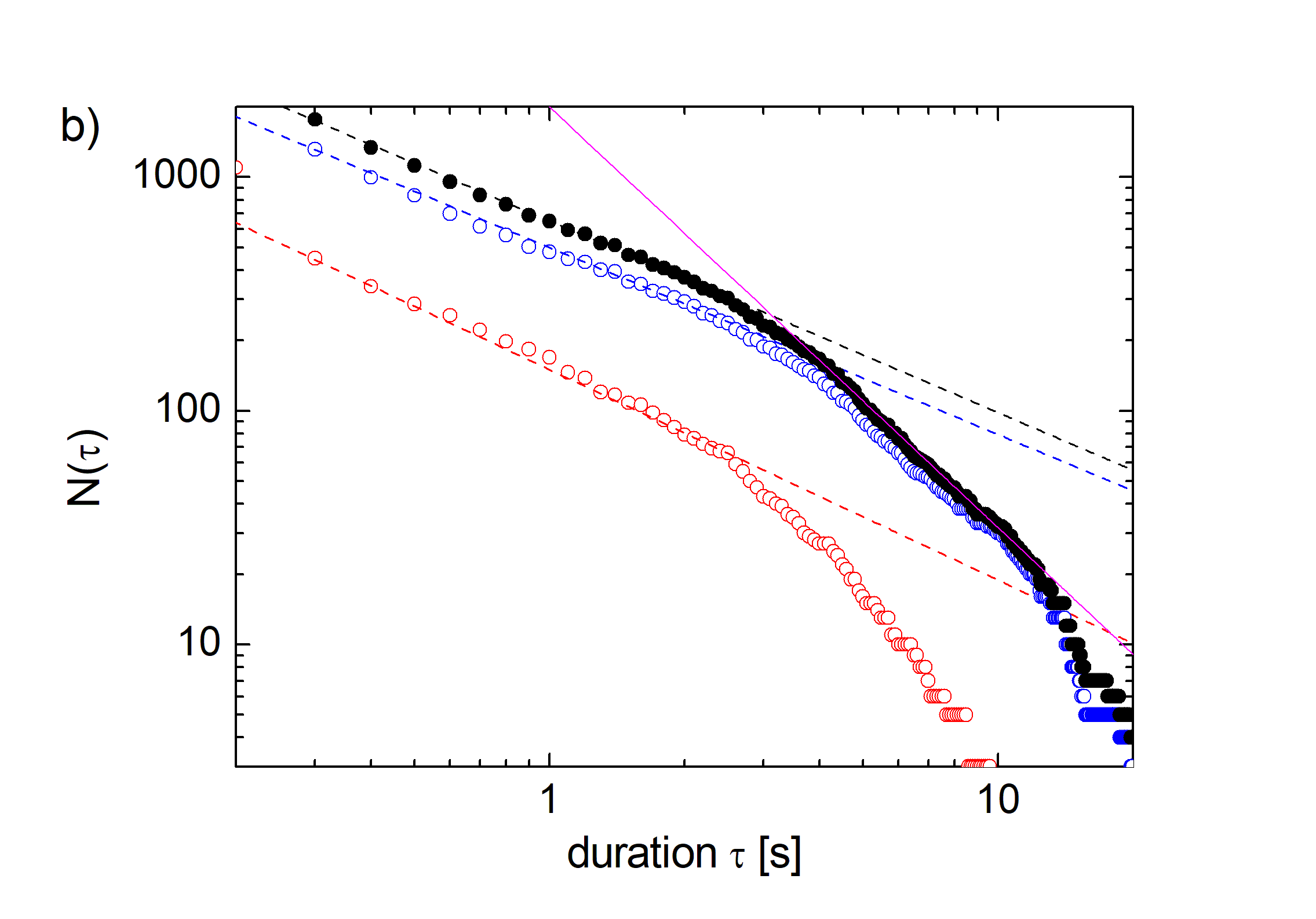}
     }
\caption{a) Number $n(\tau)$ of plateaus of durations ($\tau-0.1~{\rm s}) \dots (\tau+0.1~{\rm s}$) in the discharged mass curves, averaged over 12 discharges, 10 mm
orifice width. Red circles: clogs during the discharge of the first 400 g (approximately 1/3 of the silo content), blue circles: clogs during the remaining discharge.
b) Cumulative plots of the same data, number of clogs $N(\tau)$ of plateaus with duration longer that $\tau$. Solid symbols represent the sum of both data sets, dashed
lines reflect power laws $N(\tau)\propto\tau^{\beta}$, the solid line marks $N(\tau)\propto\tau^{\beta'}$.
}
    \label{fig:5}
\end{figure}

Figure \ref{fig:5}a presents the distribution of 'clog lengths' for the 10 mm orifice HGS-filled silo. We have evaluated the plateaus of the mass curves,
i.~e. the time intervals in which the balance readout was constant. The number $n$ of short stopped phases (up to about 3 seconds) follows approximately
a power law $n(\tau)\propto \tau^{\alpha}$ with exponent $\alpha =-1.85$. The graphs give no indication as to when the system can clearly be declared clogged.
We argue that in the athermal, passive granular system studied here, the non-permanent states of stopped outflow are never static in the complete
container. Therefore, we prefer to avoid the established technical term clog for them, and suggest the alternative denomination
'congestion' to describe the phases of stopped outflow that dissolve spontaneously in this system. They represent mere fluctuations of the discharge,
but no equilibrium states in the container. The durations of the congestions show the tendency to increase with smaller fill level, i.~e.
with decreasing pressure at the bottom of the silo, yet the exponents are practically identical. Figure \ref{fig:5}a separately shows data collected during
the outflow of the first 400 g of granular material, and during the remaining discharge until a permanently clogged state is reached.
Within the experimental uncertainty, both curves can be described with the same $\alpha$. Actually, the
exponent is reliable only for the first couple of seconds. Congestions much longer than 10 s are clearly underrepresented. This is supported in particular
by the cumulative distributions $N(\tau)$ of intervals longer than $\tau$  (see Fig.~\ref{fig:5}b). The exponent $\beta=-0.85$ of the power law
$N(\tau)\propto \tau^{\beta}$ describing the first 3 seconds is constistent with $\beta = \alpha+1$.  An exponent $\beta\ge -1$ (i.~e.~$\alpha\ge-2$) means that if the distributions would follow
these power laws for all $\tau$, the mean congestion length would diverge. For clogs longer that 10 seconds, however, the exponent $\beta'\approx -1.8$ (solid line in Fig.~\ref{fig:5}b) is well below -1 and in principle, a finite mean congestion duration exists. These quantitative scaling parameters are very preliminary, the accurate analysis of the
long-term behaviour of congestions requires much more data, particularly measurements at approximately constant fill heights.

\begin{figure}[htb]
    \centerline{
     \includegraphics[width=0.5\textwidth]{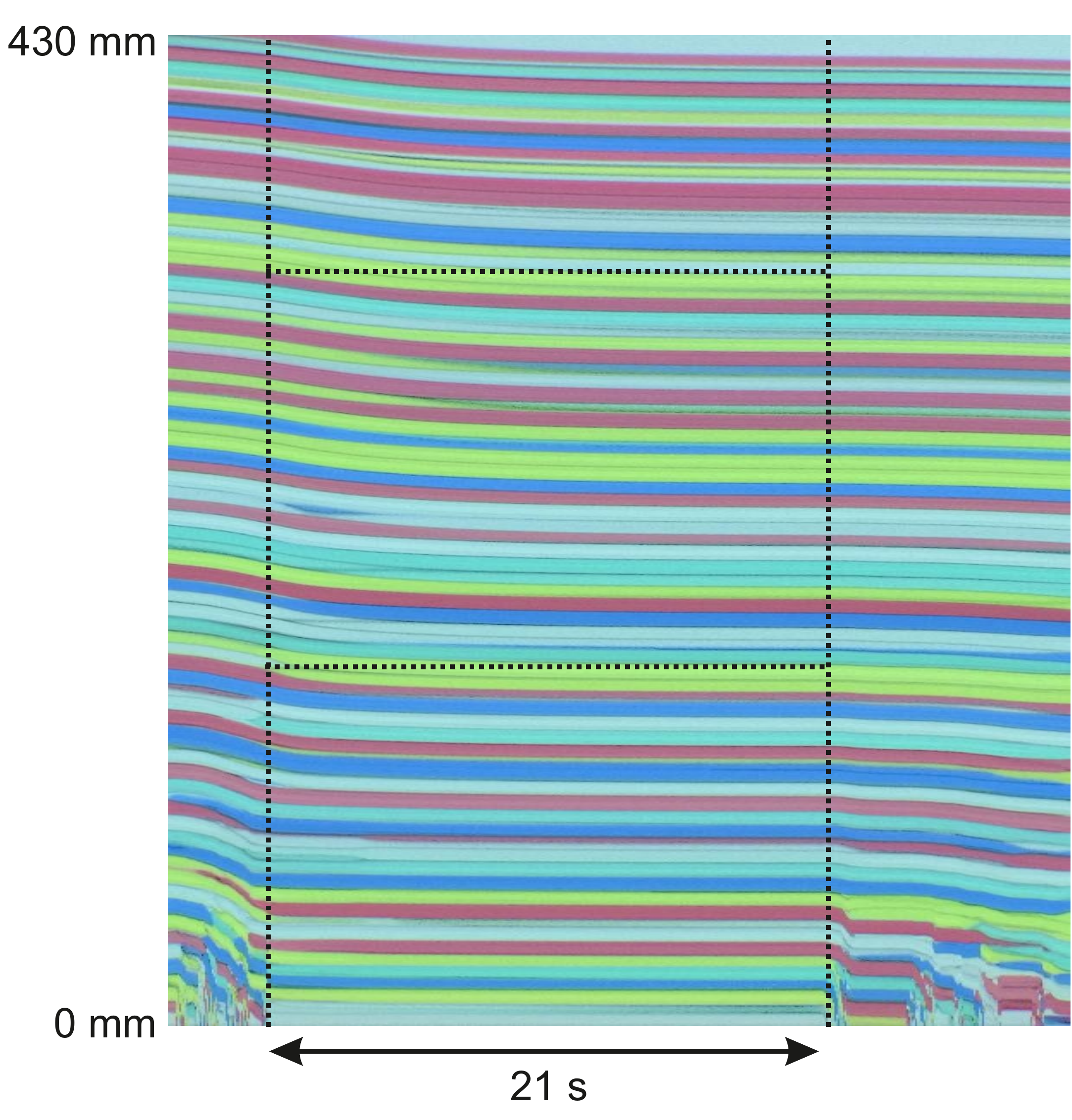}
     }
\caption{Rearrangement of the inner packing structure in the silo with 10 mm orifice shown in a space-time plot.
The interruption (same as in Fig.~\ref{fig:1}c at time 210 s), indicated by vertical dashed lines, lasts for 21 seconds.
While the particles at the bottom are immobile, one can see that the HGS in the upper layers still move down. For better visibility
of the downshifts, two dashed horizontal lines mark fixed heights.
}
    \label{fig:clog}
\end{figure}

From an external viewpoint, the observed congestions are practically indistinguishable from clogs in active and
living matter. Internally, they have a completely different origin:

After we have identified and quantitatively characterized the features of intermittent outflow of the elastic grains from the silo, we will now discuss
their physical origin: The reason for this behavior is that the material never reached an equilibrium configuration inside the silo before the blocking
arch is destroyed.
Even when the particles near the orifice locally form a stable blocking arch, there is still reorganization of the structures in the granular
bed above. These redistributions of particles and of the force network can occur anywhere in the container, and they can finally reach and affect
the blocking arch and relieve the congested state.
This is exemplarily shown in the picture of an HGS-filled silo with 10 mm opening width in Fig.~\ref{fig:clog}. The space-time plot shows a cut along
the vertical central axis of the silo, right above the orifice. Within the 35 s time window shown, the discharge stops after 3.5 s. The
position of the grains at the bottom remain practically fixed.
For the following 21 s, the outflow rate is zero. Then, the discharge sets in again. The vertical cross-sections show that the upper particles still move
downward while the blocking particles and the surroundings of the orifice remain static.

\begin{figure}[htb]
    \centerline{
     \includegraphics[width=0.32\textwidth]{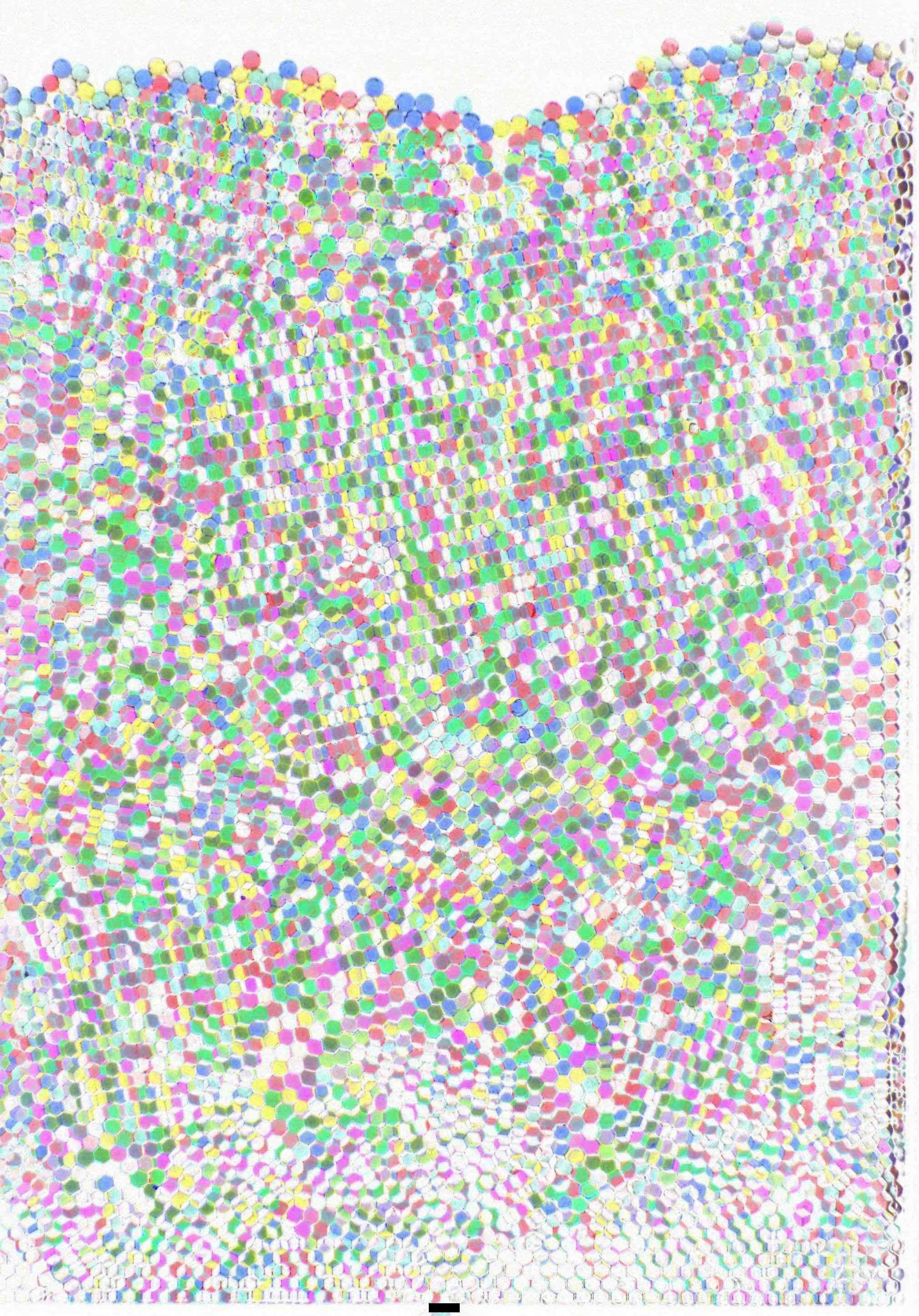}
     }
\caption{Difference image of two frames, recorded immediately after the beginning and before the end of a 21 s congested state (Fig.~\ref{fig:1}c at about 210 s,
orifice 10 mm). Image width 34 cm.
}
    \label{fig:clog2}
\end{figure}

The corresponding 2D redistributions in the complete silo are shown in Fig.~\ref{fig:clog2}: We have computed the difference between
an image recorded immediately after the outflow stopped and an image recorded shortly before the discharge continued, 21 s later.
White regions in the image are unchanged, while colors indicate where a sphere has been
replaced by one of a different color.
Near the orifice, the grains have shifted only slightly. However, one recognizes considerable reorganization and compaction in the upper parts.
The essential point is that the majority of these rearrangements proceed very slowly compared to the time scale of silo discharge (of individual grains).
This is related to viscoelasticity of the HGS. These processes set the time frame for the dissolution of blocking arches in the soft grain system.
A reasonable estimate of the time scales of such rearrangements can be extracted already from Fig. \ref{fig:dilat}b. Not only do the HGS slowly
build up pressure near the congested outlet, they can also perform quick local reconfigurations. Unless all these internal motions have ceased,
there is the chance that the force equilibrium of the blocking structure gets broken and the avalanche continues.

Permanent clogs occur only after all transient rearrangements in the container have ceased while the blocking structure is still there.
Then, the clogs are permanent and can only be destroyed by external forcing. This is regularly observed in silos with small apertures
below a certain fill-height \cite{Ashour2017b}. These observations are not specific to the quasi-2D geometry of our setup, they are
expected to be similar in 3D containers, where the observation of internal flow requires specific tomographic techniques
\cite{Stannarius2019b}.

\section{Discussion and summary}

We have analyzed the outflow of soft, low-friction hydrogel spheres from a 2D silo with small aperture experimentally, in particular
its fluctuations and intermittent character at small orifice sizes. Furthermore, the internal reorganization processes and packing
densities were monitored. We have discussed the stability of blocking structures and the life times of congestions that are specific
for the type of materials investigated here.

Fluctuations of flow rates have been described earlier for hard grains:
Unac et al. \cite{Unac2012b} reported oscillations with a narrow frequency spectrum. Such instabilities are not present
in our material. One reason may be that the HGS mass flow rate is considerably lower, and the grains are much rounder and have a viscoelastic character,
so that oscillatory density waves are rapidly dampened. Rather, we find fluctuations in a broad frequency range that are related to the slow
propagation of density and velocity fluctuations through the soft granular bed. Far regions react to local flow variations with a considerable retardation (Fig.~\ref{fig:dilat}b), so that density inhomogeneities or flow divergences are not compensated rapidly. This causes strong fluctuations in positions
and force networks,
particularly when the orifice is narrow, less than two particles wide. Then, the system can build up considerable pressure near the orifice
when the flow ceases, on a time scale of seconds. This pressure is relieved when the flow continues, causing the system to expand again.

One consequence is that the stresses that build up after the outlet is blocked and the related rearrangements of grains can spontaneously
restart the outflow from the container after some delay.
This is a qualitatively new observation for passive granular materials that are not agitated by external forcing.
Intermittent clogs that look very similar in terms of the outflow rates are found with hard particles in vibrated containers~\cite{Mankoc2009,Zuriguel2017,Guerrero2018,Guerrero2019}, in silos with oscillating bottom \cite{To2017},
or in flocks of animals passing a gate \cite{Zuriguel2014_1}. In these systems, however, momentarily existing clogs can be broken by
external forces like vibrations, or by the activity of bacteria, animals or pedestrians.
Similar intermittent clogging was recently reported in suspensions of non-Brownian microparticles \cite{Souzy2020}.
The authors suspect that these clogs are dissolved by flow through interstices of the blocking particles.

 In our passive soft sphere system, there are
no such external forces. However, the system is never in an equilibrium unless a stable overall configuration is reached. As a consequence,
the problem of the definition of avalanche sizes can be solved straightforwardly: avalanches are separated only by stable clogs that must
be destroyed by external forcing. Plateaus in the discharged mass curve $m(t)$ do not separate avalanches.
Nevertheless, for practical purposes it may be useful to define a certain empirical delay $\tau_c$ as a minimum clog duration to distinguish
avalanche events from clogs. One can then calculate mean avalanche durations and other useful statistical features.
As stated above, one has to be aware then, that this statistics will depend on the arbitrary ad hoc definition of $\tau$\ \ \cite{Cruz2018}.

In a theoretical study, Manna and Herrmann \cite{Manna2000} predicted that internal avalanches in a 2D hopper filled with frictionless
hard disks lead to self-organized criticality and intermittent clogging. In experiments, it seems that such processes are hard to observe
because the systems quickly reach an overall equilibrium or the flow continues. The time scales of internal rearrangements of the material
are comparable to those of the passage of individual particles through the orifice. In contrast, in the hydrogel system, there is a clear
separation of these time scales. Particles pass the orifice in a few hundredths of a second, but the internal redistributions of grains
require seconds. In principle, ideas similar to those simulated in Ref.~\cite{Manna2000}, cascades of internal avalanches in combination
with the elastic deformations
of the grains near the outlet may explain the intermittent outflow characteristics of soft, low-friction grains through narrow orifices.
It suggests itself that the mean time interval observed for grain rearrangements (cf. Fig.~\ref{fig:dilat}) is closely related to the
$\approx 3$~seconds range of the power law $n(\tau)$ in Figs. \ref{fig:5}a,b.

If one considers mean flow rates, then there is a clear discrepancy for small ($\rho<2$) apertures
between the behavior of the soft material studied here and Beverloo's \cite{Beverloo1961} classical equation. This arises from the fact that
Beverloo's model does not account for the clogged or congested states. Yet, even for larger orifices, the fill-level dependence of the outflow
is not explained within Beverloo's model and requires a more detailed analysis of how deformable grains pass narrow bottlenecks.

\section*{Conflicts of interest}
The authors declare no conflict of interest.

\section*{Acknowledgments}
{This project has received funding from the European Union's Horizon 2020 research and innovation programme under the Marie Sk\l{}odowska-Curie
grant agreement {\sc CALIPER} No 812638. Torsten Trittel is cordially acknowledged for important contributions to the construction of the setup.
K.H. acknowledges funding by DFG within project HA8467/2-1.
The content of this paper reflects only the authors' view and the Union is not liable for any use that may be made of the information contained therein.\smallskip

\noindent\includegraphics[width=0.09\textwidth]{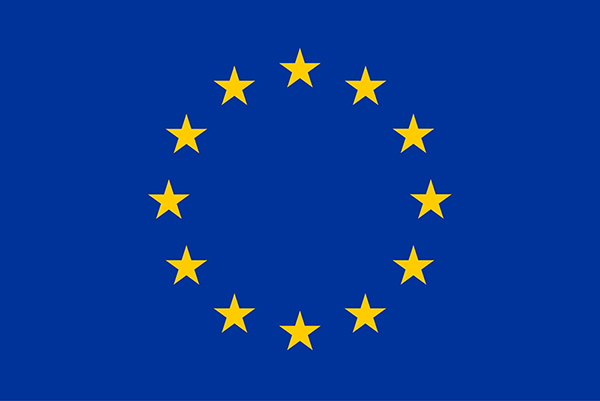}}

\bibliographystyle{plain}

\begin{mcitethebibliography}{59}
\providecommand*{\natexlab}[1]{#1}
\providecommand*{\mciteSetBstSublistMode}[1]{}
\providecommand*{\mciteSetBstMaxWidthForm}[2]{}
\providecommand*{\mciteBstWouldAddEndPuncttrue}
  {\def\EndOfBibitem{\unskip.}}
\providecommand*{\mciteBstWouldAddEndPunctfalse}
  {\let\EndOfBibitem\relax}
\providecommand*{\mciteSetBstMidEndSepPunct}[3]{}
\providecommand*{\mciteSetBstSublistLabelBeginEnd}[3]{}
\providecommand*{\EndOfBibitem}{}
\mciteSetBstSublistMode{f}
\mciteSetBstMaxWidthForm{subitem}
{(\emph{\alph{mcitesubitemcount}})}
\mciteSetBstSublistLabelBeginEnd{\mcitemaxwidthsubitemform\space}
{\relax}{\relax}

\bibitem[Huber-Burnand(1829)]{Huber-Burnand1829}
P.~Huber-Burnand, \emph{Polytechnisches Journal}, 1829, \textbf{34}, 270\relax
\mciteBstWouldAddEndPuncttrue
\mciteSetBstMidEndSepPunct{\mcitedefaultmidpunct}
{\mcitedefaultendpunct}{\mcitedefaultseppunct}\relax
\EndOfBibitem
\bibitem[Hagen(1852)]{Hagen1839}
G.~Hagen, \emph{Verhandl. der Kgl. Preu\ss{}. Akad. der Wiss. Berlin}, 1852,
  35--42\relax
\mciteBstWouldAddEndPuncttrue
\mciteSetBstMidEndSepPunct{\mcitedefaultmidpunct}
{\mcitedefaultendpunct}{\mcitedefaultseppunct}\relax
\EndOfBibitem
\bibitem[Janssen({1895})]{Janssen1895}
H.~A. Janssen, \emph{Zeitschr. d. Vereines dt. Ing.}, {1895}, \textbf{{39}},
  {1045}\relax
\mciteBstWouldAddEndPuncttrue
\mciteSetBstMidEndSepPunct{\mcitedefaultmidpunct}
{\mcitedefaultendpunct}{\mcitedefaultseppunct}\relax
\EndOfBibitem
\bibitem[Sperl({2006})]{Sperl2006}
M.~Sperl, \emph{Granular Matter}, {2006}, \textbf{{8}}, 59--65\relax
\mciteBstWouldAddEndPuncttrue
\mciteSetBstMidEndSepPunct{\mcitedefaultmidpunct}
{\mcitedefaultendpunct}{\mcitedefaultseppunct}\relax
\EndOfBibitem
\bibitem[Franklin and Johanson(1955)]{Franklin1955}
F.~C. Franklin and L.~N. Johanson, \emph{Chem. Eng. Sci.}, 1955, \textbf{4},
  119--129\relax
\mciteBstWouldAddEndPuncttrue
\mciteSetBstMidEndSepPunct{\mcitedefaultmidpunct}
{\mcitedefaultendpunct}{\mcitedefaultseppunct}\relax
\EndOfBibitem
\bibitem[Beverloo \emph{et~al.}(1961)Beverloo, Leniger, and {Van de
  Velde}]{Beverloo1961}
W.~A. Beverloo, H.~A. Leniger and J.~J. {Van de Velde}, \emph{Chem. Eng. Sci.},
  1961, \textbf{15}, 260\relax
\mciteBstWouldAddEndPuncttrue
\mciteSetBstMidEndSepPunct{\mcitedefaultmidpunct}
{\mcitedefaultendpunct}{\mcitedefaultseppunct}\relax
\EndOfBibitem
\bibitem[Neddermann \emph{et~al.}(1982)Neddermann, U.Tuzun, Savage, and
  Houlsby]{Nedderman1982}
R.~M. Neddermann, U.Tuzun, S.~B. Savage and G.~T. Houlsby, \emph{Chem. Eng.
  Sci.}, 1982, \textbf{37}, 1597--1609\relax
\mciteBstWouldAddEndPuncttrue
\mciteSetBstMidEndSepPunct{\mcitedefaultmidpunct}
{\mcitedefaultendpunct}{\mcitedefaultseppunct}\relax
\EndOfBibitem
\bibitem[Mankoc \emph{et~al.}(2007)Mankoc, Janda, Ar\'evalo, Pastor, Zuriguel,
  Garcimart\'{i}n, and D.]{Mankoc2007}
C.~Mankoc, A.~Janda, R.~Ar\'evalo, J.~M. Pastor, I.~Zuriguel,
  A.~Garcimart\'{i}n and M.~D., \emph{Granular Matter}, 2007, \textbf{9},
  407--414\relax
\mciteBstWouldAddEndPuncttrue
\mciteSetBstMidEndSepPunct{\mcitedefaultmidpunct}
{\mcitedefaultendpunct}{\mcitedefaultseppunct}\relax
\EndOfBibitem
\bibitem[Anand \emph{et~al.}(2008)Anand, Curtis, Wassgren, Hancock, and
  Ketterhagen]{Anand2008}
A.~Anand, J.~S. Curtis, C.~R. Wassgren, B.~C. Hancock and W.~R. Ketterhagen,
  \emph{Chem. Eng. Sci.}, 2008, \textbf{63}, 5821\relax
\mciteBstWouldAddEndPuncttrue
\mciteSetBstMidEndSepPunct{\mcitedefaultmidpunct}
{\mcitedefaultendpunct}{\mcitedefaultseppunct}\relax
\EndOfBibitem
\bibitem[To \emph{et~al.}(2001)To, Lai, and Pak]{To2001}
K.~To, P.~Lai and H.~K. Pak, \emph{Phys. Rev. Lett.}, 2001, \textbf{86},
  71\relax
\mciteBstWouldAddEndPuncttrue
\mciteSetBstMidEndSepPunct{\mcitedefaultmidpunct}
{\mcitedefaultendpunct}{\mcitedefaultseppunct}\relax
\EndOfBibitem
\bibitem[Hunt \emph{et~al.}(1999)Hunt, Weathers, Lee, Brennen, and
  Wassgren]{Wassgren1999}
M.~L. Hunt, R.~C. Weathers, A.~T. Lee, C.~E. Brennen and C.~R. Wassgren,
  \emph{Phys. Fluids}, 1999, \textbf{11}, 1\relax
\mciteBstWouldAddEndPuncttrue
\mciteSetBstMidEndSepPunct{\mcitedefaultmidpunct}
{\mcitedefaultendpunct}{\mcitedefaultseppunct}\relax
\EndOfBibitem
\bibitem[Wassgren \emph{et~al.}(2002)Wassgren, Hunt, Freese, Palamara, and
  Brennen]{Wassgren2002}
C.~R. Wassgren, M.~L. Hunt, P.~J. Freese, J.~Palamara and C.~E. Brennen,
  \emph{Phys. Fluids}, 2002, \textbf{14}, 10\relax
\mciteBstWouldAddEndPuncttrue
\mciteSetBstMidEndSepPunct{\mcitedefaultmidpunct}
{\mcitedefaultendpunct}{\mcitedefaultseppunct}\relax
\EndOfBibitem
\bibitem[Mankoc \emph{et~al.}(2009)Mankoc, Garcimart\'in, Zuriguel, and
  Maza]{Mankoc2009}
C.~Mankoc, A.~Garcimart\'in, I.~Zuriguel and D.~Maza, \emph{Phys. Rev. E},
  2009, \textbf{80}, 011309\relax
\mciteBstWouldAddEndPuncttrue
\mciteSetBstMidEndSepPunct{\mcitedefaultmidpunct}
{\mcitedefaultendpunct}{\mcitedefaultseppunct}\relax
\EndOfBibitem
\bibitem[Janda \emph{et~al.}(2009)Janda, Harich, Zuriguel, Maza, Cixous, and
  Garcimartin]{Janda2009}
A.~Janda, R.~Harich, I.~Zuriguel, D.~Maza, P.~Cixous and A.~Garcimartin,
  \emph{Phys. Rev. E}, 2009, \textbf{79}, 031302\relax
\mciteBstWouldAddEndPuncttrue
\mciteSetBstMidEndSepPunct{\mcitedefaultmidpunct}
{\mcitedefaultendpunct}{\mcitedefaultseppunct}\relax
\EndOfBibitem
\bibitem[Lozano \emph{et~al.}(2012)Lozano, Lumay, Zuriguel, Hidalgo, and
  A.Garcimart\'in]{Lozano2012}
C.~Lozano, G.~Lumay, I.~Zuriguel, R.~C. Hidalgo and A.Garcimart\'in,
  \emph{Phys. Rev. Lett.}, 2012, \textbf{109}, 068001\relax
\mciteBstWouldAddEndPuncttrue
\mciteSetBstMidEndSepPunct{\mcitedefaultmidpunct}
{\mcitedefaultendpunct}{\mcitedefaultseppunct}\relax
\EndOfBibitem
\bibitem[Zuriguel \emph{et~al.}(2017)Zuriguel, Janda, Ar\'evalo, Maza, and
  Garcimart\'in]{Zuriguel2017}
I.~Zuriguel, A.~Janda, R.~Ar\'evalo, D.~Maza and A.~Garcimart\'in, \emph{EPJ
  Web of Conferences}, 2017, \textbf{140}, 01002\relax
\mciteBstWouldAddEndPuncttrue
\mciteSetBstMidEndSepPunct{\mcitedefaultmidpunct}
{\mcitedefaultendpunct}{\mcitedefaultseppunct}\relax
\EndOfBibitem
\bibitem[Guerrero \emph{et~al.}(2018)Guerrero, Lozano, Zuriguel, and
  Garcimart\'in]{Guerrero2018}
B.~V. Guerrero, C.~Lozano, I.~Zuriguel and A.~Garcimart\'in, \emph{Phys. Rev.
  E}, 2018, \textbf{97}, 042904\relax
\mciteBstWouldAddEndPuncttrue
\mciteSetBstMidEndSepPunct{\mcitedefaultmidpunct}
{\mcitedefaultendpunct}{\mcitedefaultseppunct}\relax
\EndOfBibitem
\bibitem[Guerrero \emph{et~al.}(2019)Guerrero, Chakraborty, Zuriguel, and
  Garcimart\'in]{Guerrero2019}
B.~V. Guerrero, B.~Chakraborty, I.~Zuriguel and A.~Garcimart\'in, \emph{Phys.
  Rev. E}, 2019, \textbf{100}, 032901\relax
\mciteBstWouldAddEndPuncttrue
\mciteSetBstMidEndSepPunct{\mcitedefaultmidpunct}
{\mcitedefaultendpunct}{\mcitedefaultseppunct}\relax
\EndOfBibitem
\bibitem[Zuriguel \emph{et~al.}(2005)Zuriguel, Garcimart\'in, Maza, Pugnaloni,
  and Pastor]{Zuriguel2005}
I.~Zuriguel, A.~Garcimart\'in, D.~Maza, L.~A. Pugnaloni and J.~M. Pastor,
  \emph{Phys. Rev. E}, 2005, \textbf{71}, 051303\relax
\mciteBstWouldAddEndPuncttrue
\mciteSetBstMidEndSepPunct{\mcitedefaultmidpunct}
{\mcitedefaultendpunct}{\mcitedefaultseppunct}\relax
\EndOfBibitem
\bibitem[Takahashi \emph{et~al.}(1968)Takahashi, Suzuki, and
  Tanaka]{Takahashi1968}
H.~Takahashi, A.~Suzuki and T.~Tanaka, \emph{Powder Technol.}, 1968,
  \textbf{2}, 65\relax
\mciteBstWouldAddEndPuncttrue
\mciteSetBstMidEndSepPunct{\mcitedefaultmidpunct}
{\mcitedefaultendpunct}{\mcitedefaultseppunct}\relax
\EndOfBibitem
\bibitem[Suzuki \emph{et~al.}(1968)Suzuki, Takahashi, and Tanaka]{Suzuki1968}
A.~Suzuki, H.~Takahashi and T.~Tanaka, \emph{Powder Technol.}, 1968,
  \textbf{2}, 72\relax
\mciteBstWouldAddEndPuncttrue
\mciteSetBstMidEndSepPunct{\mcitedefaultmidpunct}
{\mcitedefaultendpunct}{\mcitedefaultseppunct}\relax
\EndOfBibitem
\bibitem[Lindemann and Dimon(2000)]{Lindemann2000}
K.~Lindemann and P.~Dimon, \emph{Phys. Rev. E}, 2000, \textbf{62}, 5420\relax
\mciteBstWouldAddEndPuncttrue
\mciteSetBstMidEndSepPunct{\mcitedefaultmidpunct}
{\mcitedefaultendpunct}{\mcitedefaultseppunct}\relax
\EndOfBibitem
\bibitem[Chen \emph{et~al.}(2006)Chen, Stone, Barry, Lohr, McConville, Klein,
  Sheu, Morss, Scheidemantel, and Schiffer]{Chen2006}
K.~Chen, M.~B. Stone, R.~Barry, M.~Lohr, W.~McConville, K.~Klein, B.~L. Sheu,
  A.~J. Morss, T.~Scheidemantel and P.~Schiffer, \emph{Phys. Rev. E}, 2006,
  \textbf{74}, 011306\relax
\mciteBstWouldAddEndPuncttrue
\mciteSetBstMidEndSepPunct{\mcitedefaultmidpunct}
{\mcitedefaultendpunct}{\mcitedefaultseppunct}\relax
\EndOfBibitem
\bibitem[Kumar \emph{et~al.}(2020)Kumar, Jana, Gopireddy, and Patel]{Kumar2020}
R.~Kumar, A.~K. Jana, S.~R. Gopireddy and C.~M. Patel,
  \emph{S$\bar{a}$dhan$\bar{a}$}, 2020, \textbf{45}, 67\relax
\mciteBstWouldAddEndPuncttrue
\mciteSetBstMidEndSepPunct{\mcitedefaultmidpunct}
{\mcitedefaultendpunct}{\mcitedefaultseppunct}\relax
\EndOfBibitem
\bibitem[Thomas and Durian(2015)]{Thomas2015}
C.~C. Thomas and D.~J. Durian, \emph{Phys. Rev. Lett.}, 2015, \textbf{114},
  178001\relax
\mciteBstWouldAddEndPuncttrue
\mciteSetBstMidEndSepPunct{\mcitedefaultmidpunct}
{\mcitedefaultendpunct}{\mcitedefaultseppunct}\relax
\EndOfBibitem
\bibitem[B\"orzs\"onyi \emph{et~al.}(2016)B\"orzs\"onyi, Somfai, Szab\'o,
  Wegner, Mier, Rose, and Stannarius]{Borzsonyi2016}
T.~B\"orzs\"onyi, E.~Somfai, B.~Szab\'o, S.~Wegner, P.~Mier, G.~Rose and
  R.~Stannarius, \emph{New J. Phys.}, 2016, \textbf{18}, 093017\relax
\mciteBstWouldAddEndPuncttrue
\mciteSetBstMidEndSepPunct{\mcitedefaultmidpunct}
{\mcitedefaultendpunct}{\mcitedefaultseppunct}\relax
\EndOfBibitem
\bibitem[Ashour \emph{et~al.}(2017)Ashour, Wegner, Trittel, B\"orzs\"onyi, and
  Stannarius]{Ashour2017}
A.~Ashour, S.~Wegner, T.~Trittel, T.~B\"orzs\"onyi and R.~Stannarius,
  \emph{Soft Matter}, 2017, \textbf{13}, 402\relax
\mciteBstWouldAddEndPuncttrue
\mciteSetBstMidEndSepPunct{\mcitedefaultmidpunct}
{\mcitedefaultendpunct}{\mcitedefaultseppunct}\relax
\EndOfBibitem
\bibitem[Szab\'o \emph{et~al.}(2018)Szab\'o, Kov\'acs, Wegner, Ashour, Fischer,
  Stannarius, and B\"orzs\"onyi]{Szabo2018}
B.~Szab\'o, Z.~Kov\'acs, S.~Wegner, A.~Ashour, D.~Fischer, R.~Stannarius and
  T.~B\"orzs\"onyi, \emph{Phys. Rev. E}, 2018, \textbf{97}, 062904\relax
\mciteBstWouldAddEndPuncttrue
\mciteSetBstMidEndSepPunct{\mcitedefaultmidpunct}
{\mcitedefaultendpunct}{\mcitedefaultseppunct}\relax
\EndOfBibitem
\bibitem[T\"or\"ok \emph{et~al.}(2017)T\"or\"ok, L\'evay, Szab\'o, Somfai,
  Wegner, Stannarius, and B\"orzs\"onyi]{Torok2017}
J.~T\"or\"ok, S.~L\'evay, B.~Szab\'o, E.~Somfai, S.~Wegner, R.~Stannarius and
  T.~B\"orzs\"onyi, \emph{EPJ Web of Conferences}, 2017, \textbf{140},
  03076\relax
\mciteBstWouldAddEndPuncttrue
\mciteSetBstMidEndSepPunct{\mcitedefaultmidpunct}
{\mcitedefaultendpunct}{\mcitedefaultseppunct}\relax
\EndOfBibitem
\bibitem[Lattanzi and Stickel(2019)]{Lattanzi2019}
A.~M. Lattanzi and J.~J. Stickel, \emph{AICHE Journal}, 2019, \textbf{66},
  e16882\relax
\mciteBstWouldAddEndPuncttrue
\mciteSetBstMidEndSepPunct{\mcitedefaultmidpunct}
{\mcitedefaultendpunct}{\mcitedefaultseppunct}\relax
\EndOfBibitem
\bibitem[Tang and Behringer(2011)]{Tang2011}
J.~Tang and R.~Behringer, \emph{Chaos}, 2011, \textbf{21}, 041107\relax
\mciteBstWouldAddEndPuncttrue
\mciteSetBstMidEndSepPunct{\mcitedefaultmidpunct}
{\mcitedefaultendpunct}{\mcitedefaultseppunct}\relax
\EndOfBibitem
\bibitem[Zuriguel \emph{et~al.}(2014)Zuriguel, Parisi, Hidalgo, Lozano, Janda,
  Gago, Peralta, Ferrer, Pugnaloni, Cl\'ement, Maza, Pagonabarraga, and
  Garcimart\'in]{Zuriguel2014_1}
I.~Zuriguel, D.~R. Parisi, R.~C. Hidalgo, C.~Lozano, A.~Janda, P.~A. Gago,
  J.~P. Peralta, L.~M. Ferrer, L.~A. Pugnaloni, E.~Cl\'ement, D.~Maza,
  I.~Pagonabarraga and A.~Garcimart\'in, \emph{Scientific Reports}, 2014,
  \textbf{4}, 7324\relax
\mciteBstWouldAddEndPuncttrue
\mciteSetBstMidEndSepPunct{\mcitedefaultmidpunct}
{\mcitedefaultendpunct}{\mcitedefaultseppunct}\relax
\EndOfBibitem
\bibitem[Rubio-Largo \emph{et~al.}(2015)Rubio-Largo, Janda, Maza, Zuriguel, and
  Hidalgo]{Rubio-largo2015}
S.~M. Rubio-Largo, A.~Janda, D.~Maza, I.~Zuriguel and R.~C. Hidalgo,
  \emph{Phys. Rev. Lett.}, 2015, \textbf{114}, 238002\relax
\mciteBstWouldAddEndPuncttrue
\mciteSetBstMidEndSepPunct{\mcitedefaultmidpunct}
{\mcitedefaultendpunct}{\mcitedefaultseppunct}\relax
\EndOfBibitem
\bibitem[Hidalgo \emph{et~al.}(2013)Hidalgo, Lozano, Zuriguel, and
  Garcimart\'in]{Hidalgo2013}
R.~C. Hidalgo, C.~Lozano, I.~Zuriguel and A.~Garcimart\'in, \emph{Granular
  Matter}, 2013, \textbf{15}, 841--848\relax
\mciteBstWouldAddEndPuncttrue
\mciteSetBstMidEndSepPunct{\mcitedefaultmidpunct}
{\mcitedefaultendpunct}{\mcitedefaultseppunct}\relax
\EndOfBibitem
\bibitem[Vivanco \emph{et~al.}(2012)Vivanco, Rica, and Melo]{Vivanco2012}
F.~Vivanco, S.~Rica and F.~Melo, \emph{Granular Matter}, 2012, \textbf{14},
  563--576\relax
\mciteBstWouldAddEndPuncttrue
\mciteSetBstMidEndSepPunct{\mcitedefaultmidpunct}
{\mcitedefaultendpunct}{\mcitedefaultseppunct}\relax
\EndOfBibitem
\bibitem[Drescher \emph{et~al.}(1995)Drescher, Waters, and
  Rhoades]{Drescher1995I}
A.~Drescher, A.~J. Waters and C.~A. Rhoades, \emph{Powder Technol.}, 1995,
  \textbf{84}, 165\relax
\mciteBstWouldAddEndPuncttrue
\mciteSetBstMidEndSepPunct{\mcitedefaultmidpunct}
{\mcitedefaultendpunct}{\mcitedefaultseppunct}\relax
\EndOfBibitem
\bibitem[Drescher \emph{et~al.}(1995)Drescher, Waters, and
  Rhoades]{Drescher1995II}
A.~Drescher, A.~J. Waters and C.~A. Rhoades, \emph{Powder Technol.}, 1995,
  \textbf{84}, 177\relax
\mciteBstWouldAddEndPuncttrue
\mciteSetBstMidEndSepPunct{\mcitedefaultmidpunct}
{\mcitedefaultendpunct}{\mcitedefaultseppunct}\relax
\EndOfBibitem
\bibitem[Manna and Herrmann(2000)]{Manna2000}
S.~S. Manna and H.~J. Herrmann, \emph{Eur. Phys. J. E}, 2000, \textbf{1},
  341--344\relax
\mciteBstWouldAddEndPuncttrue
\mciteSetBstMidEndSepPunct{\mcitedefaultmidpunct}
{\mcitedefaultendpunct}{\mcitedefaultseppunct}\relax
\EndOfBibitem
\bibitem[Pugnaloni \emph{et~al.}(2001)Pugnaloni, Barker, and
  Mehta]{Pugnaloni2001}
L.~A. Pugnaloni, G.~C. Barker and A.~Mehta, \emph{Adv. Complex Syst.}, 2001,
  \textbf{4}, 289\relax
\mciteBstWouldAddEndPuncttrue
\mciteSetBstMidEndSepPunct{\mcitedefaultmidpunct}
{\mcitedefaultendpunct}{\mcitedefaultseppunct}\relax
\EndOfBibitem
\bibitem[To and Lai(2002)]{To2002}
K.~To and P.-Y. Lai, \emph{Phys. Rev. E}, 2002, \textbf{66}, 011308\relax
\mciteBstWouldAddEndPuncttrue
\mciteSetBstMidEndSepPunct{\mcitedefaultmidpunct}
{\mcitedefaultendpunct}{\mcitedefaultseppunct}\relax
\EndOfBibitem
\bibitem[Zuriguel \emph{et~al.}(2003)Zuriguel, Pugnaloni, Garcimart\'in, and
  Maza]{Zuriguel2003}
I.~Zuriguel, L.~A. Pugnaloni, A.~Garcimart\'in and D.~Maza, \emph{Phys. Rev.
  E}, 2003, \textbf{68}, 030301\relax
\mciteBstWouldAddEndPuncttrue
\mciteSetBstMidEndSepPunct{\mcitedefaultmidpunct}
{\mcitedefaultendpunct}{\mcitedefaultseppunct}\relax
\EndOfBibitem
\bibitem[Pugnaloni and Barker(2004)]{Pugnaloni2004}
L.~A. Pugnaloni and G.~C. Barker, \emph{Physica A}, 2004, \textbf{337},
  428\relax
\mciteBstWouldAddEndPuncttrue
\mciteSetBstMidEndSepPunct{\mcitedefaultmidpunct}
{\mcitedefaultendpunct}{\mcitedefaultseppunct}\relax
\EndOfBibitem
\bibitem[Garcimart\'in \emph{et~al.}(2013)Garcimart\'in, Lozano, Lumay, and
  Zuriguel]{Garcimartin2013}
A.~Garcimart\'in, C.~Lozano, G.~Lumay and I.~Zuriguel, \emph{AIP Conference
  Proceedings}, 2013, \textbf{1542}, 686\relax
\mciteBstWouldAddEndPuncttrue
\mciteSetBstMidEndSepPunct{\mcitedefaultmidpunct}
{\mcitedefaultendpunct}{\mcitedefaultseppunct}\relax
\EndOfBibitem
\bibitem[Marin \emph{et~al.}(2018)Marin, Lhuissier, Rossi, and
  K\"ahler]{Marin2018}
A.~Marin, H.~Lhuissier, M.~Rossi and C.~J. K\"ahler, \emph{Phys. Rev. E}, 2018,
  \textbf{97}, 021102(R)\relax
\mciteBstWouldAddEndPuncttrue
\mciteSetBstMidEndSepPunct{\mcitedefaultmidpunct}
{\mcitedefaultendpunct}{\mcitedefaultseppunct}\relax
\EndOfBibitem
\bibitem[Souzy \emph{et~al.}(2020)Souzy, Zuriguel, and Marin]{Souzy2020}
M.~Souzy, I.~Zuriguel and A.~Marin, \emph{arxiv.org: 2004.04413}, 2020\relax
\mciteBstWouldAddEndPuncttrue
\mciteSetBstMidEndSepPunct{\mcitedefaultmidpunct}
{\mcitedefaultendpunct}{\mcitedefaultseppunct}\relax
\EndOfBibitem
\bibitem[Hong \emph{et~al.}(2017)Hong, Kohne, Morrell, Wang, and
  Weeks]{Hong2017}
X.~Hong, M.~Kohne, M.~Morrell, H.~Wang and E.~R. Weeks, \emph{Phys. Rev. E},
  2017, \textbf{96}, 062605\relax
\mciteBstWouldAddEndPuncttrue
\mciteSetBstMidEndSepPunct{\mcitedefaultmidpunct}
{\mcitedefaultendpunct}{\mcitedefaultseppunct}\relax
\EndOfBibitem
\bibitem[Ashour \emph{et~al.}(2017)Ashour, Trittel, B\"orzs\"onyi, and
  Stannarius]{Ashour2017b}
A.~Ashour, T.~Trittel, T.~B\"orzs\"onyi and R.~Stannarius, \emph{Phys. Rev.
  Fluids}, 2017, \textbf{2}, 123302\relax
\mciteBstWouldAddEndPuncttrue
\mciteSetBstMidEndSepPunct{\mcitedefaultmidpunct}
{\mcitedefaultendpunct}{\mcitedefaultseppunct}\relax
\EndOfBibitem
\bibitem[Stannarius \emph{et~al.}(2019)Stannarius, Martinez, Finger, Somfai,
  and B\"orzs\"onyi]{Stannarius2019}
R.~Stannarius, D.~S. Martinez, T.~Finger, E.~Somfai and T.~B\"orzs\"onyi,
  \emph{Granular Matter}, 2019, \textbf{21}, 56\relax
\mciteBstWouldAddEndPuncttrue
\mciteSetBstMidEndSepPunct{\mcitedefaultmidpunct}
{\mcitedefaultendpunct}{\mcitedefaultseppunct}\relax
\EndOfBibitem
\bibitem[Stannarius \emph{et~al.}(2019)Stannarius, Martinez, B\"orzs\"ony,
  Bieberle, Barthel, and Hampel]{Stannarius2019b}
R.~Stannarius, D.~S. Martinez, T.~B\"orzs\"ony, M.~Bieberle, F.~Barthel and
  U.~Hampel, \emph{New J. Phys.}, 2019, \textbf{21}, 113054\relax
\mciteBstWouldAddEndPuncttrue
\mciteSetBstMidEndSepPunct{\mcitedefaultmidpunct}
{\mcitedefaultendpunct}{\mcitedefaultseppunct}\relax
\EndOfBibitem
\bibitem[Bertho \emph{et~al.}(2006)Bertho, Becco, and Vandewalle]{Bertho2006}
Y.~Bertho, C.~Becco and N.~Vandewalle, \emph{Phys. Rev. E}, 2006, \textbf{73},
  056309\relax
\mciteBstWouldAddEndPuncttrue
\mciteSetBstMidEndSepPunct{\mcitedefaultmidpunct}
{\mcitedefaultendpunct}{\mcitedefaultseppunct}\relax
\EndOfBibitem
\bibitem[Hong \emph{et~al.}(2017)Hong, Kohne, and Weeks]{Hong2017b}
X.~Hong, M.~Kohne and E.~R. Weeks, \emph{arXiv: 1512.02500v2}, 2017\relax
\mciteBstWouldAddEndPuncttrue
\mciteSetBstMidEndSepPunct{\mcitedefaultmidpunct}
{\mcitedefaultendpunct}{\mcitedefaultseppunct}\relax
\EndOfBibitem
\bibitem[Zuriguel(2014)]{Zuriguel2014}
I.~Zuriguel, \emph{Papers in Physics}, 2014, \textbf{6}, 060014\relax
\mciteBstWouldAddEndPuncttrue
\mciteSetBstMidEndSepPunct{\mcitedefaultmidpunct}
{\mcitedefaultendpunct}{\mcitedefaultseppunct}\relax
\EndOfBibitem
\bibitem[Helbing and Molnar(1995)]{Helbing1995}
D.~Helbing and P.~Molnar, \emph{Phys. Rev. E}, 1995, \textbf{51},
  4282--4286\relax
\mciteBstWouldAddEndPuncttrue
\mciteSetBstMidEndSepPunct{\mcitedefaultmidpunct}
{\mcitedefaultendpunct}{\mcitedefaultseppunct}\relax
\EndOfBibitem
\bibitem[Helbing \emph{et~al.}(2000)Helbing, Farkas, and Vicsek]{Helbing2000b}
D.~Helbing, I.~J. Farkas and T.~Vicsek, \emph{Nature}, 2000, \textbf{407},
  487--490\relax
\mciteBstWouldAddEndPuncttrue
\mciteSetBstMidEndSepPunct{\mcitedefaultmidpunct}
{\mcitedefaultendpunct}{\mcitedefaultseppunct}\relax
\EndOfBibitem
\bibitem[Helbing \emph{et~al.}(2005)Helbing, Buzna, Johansson, and
  Werner]{Helbing2005}
D.~Helbing, L.~Buzna, A.~Johansson and T.~Werner, \emph{Transportation
  Science}, 2005, \textbf{39}, 1--24\relax
\mciteBstWouldAddEndPuncttrue
\mciteSetBstMidEndSepPunct{\mcitedefaultmidpunct}
{\mcitedefaultendpunct}{\mcitedefaultseppunct}\relax
\EndOfBibitem
\bibitem[SI()]{SI}
{A video with a 60 s sequence from the discharge of HGS from a 2D silo with 10
  mm orifice width is provided as supplemental material.}\relax
\mciteBstWouldAddEndPunctfalse
\mciteSetBstMidEndSepPunct{\mcitedefaultmidpunct}
{}{\mcitedefaultseppunct}\relax
\EndOfBibitem
\bibitem[Hidalgo \emph{et~al.}(2018)Hidalgo, Go$\tilde{n}$i-Arana,
  Hern\'andez-Puerta, and Pagonabarraga]{Cruz2018}
R.~C. Hidalgo, A.~Go$\tilde{n}$i-Arana, A.~Hern\'andez-Puerta and
  I.~Pagonabarraga, \emph{Phys. Rev. E}, 2018, \textbf{97}, 012611\relax
\mciteBstWouldAddEndPuncttrue
\mciteSetBstMidEndSepPunct{\mcitedefaultmidpunct}
{\mcitedefaultendpunct}{\mcitedefaultseppunct}\relax
\EndOfBibitem
\bibitem[Unac \emph{et~al.}(2012)Unac, Vidales, Benegas, and
  Ippolito]{Unac2012b}
R.~O. Unac, A.~M. Vidales, O.~A. Benegas and I.~Ippolito, \emph{Powder
  Technol.}, 2012, \textbf{225}, 214--220\relax
\mciteBstWouldAddEndPuncttrue
\mciteSetBstMidEndSepPunct{\mcitedefaultmidpunct}
{\mcitedefaultendpunct}{\mcitedefaultseppunct}\relax
\EndOfBibitem
\bibitem[To and Tai(2017)]{To2017}
K.~To and H.-T. Tai, \emph{Phys. Rev. E}, 2017, \textbf{96}, 032906\relax
\mciteBstWouldAddEndPuncttrue
\mciteSetBstMidEndSepPunct{\mcitedefaultmidpunct}
{\mcitedefaultendpunct}{\mcitedefaultseppunct}\relax
\EndOfBibitem
\end{mcitethebibliography}
\providecommand*{\mcitethebibliography}{\thebibliography}
\csname @ifundefined\endcsname{endmcitethebibliography}
{\let\endmcitethebibliography\endthebibliography}{}

\end{document}